\documentclass[12pt, pra, onecolumn, showpacs, amsmath,amssymb, superscriptaddress, twoside]{revtex4-1}

\usepackage{amssymb}

\usepackage{txfonts}

\usepackage{empheq}

\usepackage{graphicx}
\usepackage{dcolumn}
\usepackage{bm}

\usepackage{CJKutf8}

\begin{document}
\begin{CJK}{UTF8}{gbsn}

\title{Direct manipulation of wave amplitude and phase \\through inverse design of isotropic media}

\date{accepted 7 April, proofread 14 June, compiled \today}

\author{Yangji\'e Liu (刘\CJKkern泱\CJKkern杰)}
\email[Electronic address:] {yangjie@hubu.edu.cn}

\affiliation{Antennas Group, School of Electrical Engineering and Computer Science, Queen Mary University of London, 327 Mile End Road, E1 4NS, England, United Kingdom}
\affiliation{Faculty of Physics and Electronic Science, Hubei University, 368 Youyi Road, 430062 Wuchang District, Wuhan P. R. China}

\author{B. Vial}
\affiliation{Antennas Group, School of Electrical Engineering and Computer Science, Queen Mary University of London, 327 Mile End Road, E1 4NS, England, United Kingdom}

\author{S. A. R. Horsley}
\affiliation{Department of Physics and Astronomy, University of Exeter, Stocker Road, Exeter, EX4~4QL, England, United Kingdom}

\author{T. G. Philbin}
\affiliation{Department of Physics and Astronomy, University of Exeter, Stocker Road, Exeter, EX4~4QL, England, United Kingdom}

\author{Y. Hao}
\affiliation{Antennas Group, School of Electrical Engineering and Computer Science, Queen Mary University of London, 327 Mile End Road, E1 4NS, England, United Kingdom}


\begin{abstract}

\noindent In this article we propose a new design methodology allowing us to control both amplitude and phase of electromagnetic waves from a cylindrical incident wave. This results in isotropic materials and does not resort to transformation optics or its quasi-conformal approximations. Our method leads to two-dimensional isotropic, inhomogeneous material profiles of permittivity and permeability, to which a general class of scattering-free wave solutions arise. Our design is based on the separation of the complex wave solution into amplitude and phase. We give two types of examples to validate our methodology. \footnote{Submitted to \emph{New J. Phys.} 28 Dec-2016, initial decision received 23 Feb, deposited 21 Jan in arXiv (\url{https://arxiv.org/abs/1701.06021v1}), NJP-106316.R1 resubmitted 14 March, accepted 7 April 2017 as Manuscript \# aa6c0c (\url{http://iopscience.iop.org/article/10.1088/1367-2630/aa6c0c/}), 6 July published online, 3 Aug 2017 updated in arXiv (\url{https://arxiv.org/abs/1701.06021v2}). }

\end{abstract}


\maketitle

\end{CJK}

\section{Introduction}

Transformation optics (TO) has proved as a remarkable design tool to engineer materials which are capable of manipulating waves in an arbitrary manner~\cite{Pendry2006, Leonhardt2006a, Leonhardt2006b, Leonhardt2010book, Post1962}. It yields material parameters when wave behaviours are predefined by coordinate deformation, the wave remaining impedance-matched to the virtual space (usually the air). However, this method or its derivatives such as quasi-conformal mapping are limited to problems with a ``mother" design required in a virtual space. What remains less explored is the vast space of inhomogeneous isotropic media~\cite{Leonhardt2010book,Schmiele2010}, especially those not of transformation media. While the material parameters derived from transformation optics are generally anisotropic, the quasi-conformal transformation can be applied to reduce the anisotropicity~\cite{Li2008} and other numerical techniques can also be applied to achieve other types of inverse isotropic design~\cite{Chang2010, LiuD2013, Piggott2015, Ginis2016}. Thus the material parameters shall be more readily manufactured if we can design isotropic materials without resorting to quasi-conformal approximations~\cite{Li2008}. 

Generally a different approach to design the isotropic, graded index medium for light is demanded\cite{Leonhardt2009c,Leonhardt2010book}. We are thus tempted to look into the analytic relation between the wave solution and material parameters, in the hope of gaining some new insight about the design philosophy of tackling the optical inverse problem in inhomogeneous media. This approach will generally result in 2D graded index profiles of media, other than only 1D graded indices of refraction~\cite{Dalarsson2009}.

In this paper we will demonstrate a method for designing graded-index, isotropic material profiles of an optical inverse problem which give scattering-free wave solutions~\cite{Ben2016b}. We will start from the wave equation~\cite{Philbin2014} and show the otherwise implicit relation between a reflection-free wave solution in polar form and its corresponding material profiles. This is an \emph{alternative} method to compute material parameters other than transformation optics, and circumvents the trail-and-error way to guess parameters.

We emphasise that our theory is a general approach to designing material profiles for non-reflective electromagnetic waves, and should be extendable to vectorial, three-dimensional cases despite the mathematical complication. For simple illustration, we restrain our investigation in this paper to two-dimensional case with linear polarisations~\cite{Kim2013}. The merit of the analysis applied here lies in its simplicity to design the material profile without resorting to numerics. A remark to note is that our design profiles of material are real-valued, which surprisingly supports reflection-less waves for certain incident waves. This contrasts with the complex material profiles that work for a single angle of incidence~\footnote{Taking $E_z=F(x,y)\exp(i k_0 x)$ with $F\rightarrow1$ at infinity and plugging this into the wave equation of no source~\eqref{Ez}, one reads off the corresponding permittivity. }.

The structure of the paper is as follow. The methodology for designing material profiles to achieve predefined amplitude or phase will be given in Sec.~\ref{meth}. The example material profiles in this paper are designed for two types of wave solutions: 1) a spatial-modulating wave from a point source (Sec.~\ref{Sec3}), which keeps its phase as if in free space; 2) a wave transiting from a point source to another phase (Sec.~\ref{converter}). Therefore we demonstrate how one can control the wave front -- maintaining it or varying it at our will, respectively. Sec.~\ref{Ego} remarks about the condition when the wave equation Eq.~\eqref{Ez} aligns with exact geometrical optics~\cite{Philbin2014}. Sec.~\ref{conc} concludes the article.

\section{Methodology}\label{meth}
We start from vectorial Maxwell equations for 2D isotropic, inhomogeneous and passive medium, where the relative permittivity and permeability $\epsilon,\mu$ are real, spatial-varying functions~\cite{BornWolf1999}. We aim to design the material profiles $\epsilon,\mu$ which both reduce to unity in the far region ($\sqrt{x^2+y^2}\sim \infty$). The electromagnetic wave in two-dimensional case, can be decoupled into two separate scalar solutions: transverse electric (TE) and transverse magnetic (TM) components. For TE component, we represent its scalar parts in frequency domain, in polar form of real amplitude and phase as ~\footnote{To be compatible with the model setup in \texttt{COMSOL} simulation, we choose the engineering time harmonic convection $e^{i\omega t}$.}: 
\begin{eqnarray}
\label{singleE}
\bold E=\hat{z}E_z=\hat{z}A e^{+i\omega t-i\phi}, 
\end{eqnarray}
where the TE waves $E_z$ is composed of amplitudes $A$ and phase $\phi$. Explicitly the TE wave equation is written as 
\begin{eqnarray}
\label{Ez}
\nabla^2 E_z+\epsilon\mu k_0^2 E_z-\frac{1}{\mu}\nabla \mu\cdot\nabla E_z=0,
\end{eqnarray}
where $k_0=\omega/c$ with $c$ the light velocity in vacuum. Eq.~\eqref{Ez} will be the main wave equation we use in this article. The magnetic field for TE mode is given in terms of the electric field by Faraday's law of induction, when the electric field is known:
\begin{equation}
\bold H=i\frac{\nabla\times E_z\hat{z}}{\mu_0\mu\omega}.
\end{equation}
As for TM polarisation and the solution for $\mathbf{E}$ fileds, similar results hold when the permittivity and permeability are swapped. 

In this article, we will focus on the TE wave in Eq.~\eqref{Ez}. From Eqs.~\eqref{singleE} and \eqref{Ez} we obtain the equivalent of the wave equation in terms of amplitude and phase
 \begin{empheq}[left=\empheqlbrace]{align}
&\nabla^2 A-A\nabla \phi\cdot \nabla \phi+\epsilon\mu k_0^2 A-\frac{1}{\mu}\nabla\mu\cdot \nabla A=0,   \label{mains1} \\
&\nabla\cdot \big(\frac{A^2}{\mu}\nabla \phi\big)=0.   \label{mains2}
  \end{empheq}

We now analyse wave manipulation when it traverses the inhomogeneous medium. The material profiles are always designed to reduce to one in the far region. We choose an incident wave vector $k_0\hat{n}$, in which $\hat{n}$ is its unit vector. We interpret the wave dynamics as a system where an incidence wave from the source impinges the whole medium $\epsilon(x, y), \mu(x, y)$ from one side, then interacts with the non-trivial graded index medium in the middle. Finally it exits to the other side the medium (air) with no scattering. The gradient of the phase is written as a sum of incident wave vector $\nabla \phi=k_0\hat{n}$ and the gradient of an unknown function $\psi(x, y)$:
\begin{eqnarray}
\label{phase}
\nabla \phi=k_0\hat{n}+\nabla \psi, 
\end{eqnarray}
Then Eq.~\eqref{mains2} can be written as Poisson's equation, with $\nabla\psi$ playing the role of the scalar potential. This can be solved in the same way as the equivalent problem in 2D electrostatics, where $\bold {E}={1}/{(2\pi\epsilon_0)}\int{\rm d}^2\bold{x}'\rho(\bold{x}'){(\bold{x}-\bold{x'})}/{\vert\bold{x}-\bold{x'}\vert^2}$(cf. pp4-7\cite{YangjieThesis}). Therefore $\nabla \psi$ can be given as
\begin{eqnarray}
\label{newnablas}
\nabla \psi(\mathbf{x})=-\frac{k_0\mu(\mathbf{x})}{2\pi A^2(\mathbf{x})}\int{\rm d}^2\mathbf{x'}\frac{\mathbf{x}-\mathbf{x'}}{\vert\mathbf{x}-\mathbf{x'}\vert^2}\nabla'\cdot\Big[{\frac{A^2(\mathbf{x'})}{\mu(\mathbf{x'})}\hat{n}(\mathbf{x'})}\Big],
\end{eqnarray}
where $\mathbf{x}:=\hat{x}x+\hat{y}y, \mathbf{x}':=\hat{x}x'+\hat{y}y'$. This integral physically indicates that the incident wave vector and spatially dependent amplitude $\nabla\cdot({A^2}\hat{n}/\mu)$ serve as the source of $\nabla \psi$.

An interesting case occurs when $\nabla \psi$ vanishes everywhere, \emph{i.e.} the wave not only exits the medium as the same incident wave, but also keeps its phase unchanged throughout space with respect to the incident wave, even inside the medium. This occurs when the source term $\nabla\cdot(A^2\hat{n}/\mu)$ vanishes in Eq.~\eqref{newnablas}. We shall term the medium an invisible material for a certain incidence wave and its examples will shown in Sec.~\ref{Sec3}. In another case when $\nabla \psi$ does \emph{not} vanish throughout the medium, the incidence wave shall withstand a non-zero phase change in the middle of medium, and then outputs as a different wave. This medium shapes one incidence wave into another. We shall show in Sec.~\ref{converter} two instances of such a type of conversion medium, which transforms a cylindrical wave into others.

\subsection{Controlling the wave amplitude}\label{method1}
Now we demonstrate our method to design a medium, 
which controls the wave amplitude $A$ generated from a localised current source $j(x,y;\omega)$ in free space and leaves the wave phase unchanged as that in free space. 

Generally in a graded index medium $(\epsilon, \mu)$ the wave equation with source $j(x,y;\omega)$ in frequency domain is
\begin{equation}\label{Ezgen}
\nabla^2 E_z+k_0^2 \epsilon\mu E_z-\frac{\nabla\mu}{\mu}\cdot \nabla E_z = i\omega\mu_0\mu j, 
\end{equation}
similar to Eq.~\eqref{Ez}. In free space air, Eq.~\eqref{Ezgen} reduces to 
\begin{equation}\label{refE0}
\nabla^2 E_z^0+k_0^2 E_z^0 = i\omega\mu_0j, 
\end{equation}
in which $E_z^0=A_0 e^{-iS_0}$ is the reference wave solution resulting from a source $j(x, y; \omega)$. The form of Eqs.~\eqref{Ezgen} and \eqref{refE0} includes the special case of planar wave of incidence, i.e. the source term $j$ is infinite away and thus equivalent to zero.

Supposing the source is positioned in a bounded region in air, the source-free wave equation Eq.~\eqref{Ez} actually still stands all over the field region \emph{except in the source region}. This means that we can disregard the source term in Eqs.~\eqref{Ezgen} and \eqref{refE0}, and still use Eqs.~\eqref{mains1} and~\eqref{mains2} derived from Eq.~\eqref{Ez}~\footnote{The general situation when the source sits in non-trivial region where material parameters are not unity, is far more complicated and beyond scope of this article. }. We want to control the wave amplitude $A$ as
\begin{equation}
\label{amp}
A:=fA_0, 
\end{equation}
where $f(x, y)$ is the modulation factor determined by the designer with respect to the reference wave solution, i.e. $A_0$. The modulation factor $f$ can be arbitrary function as long as it reduces to unity in far field. Then substituting Eq.~\eqref{refE0} into Eqs.~\eqref{mains1}-\eqref{mains2} we have in source-free region
\begin{empheq}[left=\empheqlbrace]{align}
\label{fG}
&\nabla^2 (fA_0)-fA_0\nabla S_0\cdot \nabla S_0+\nonumber\\
&\epsilon\mu k_0^2 fA_0-
\frac{\nabla\mu}{\mu}\cdot\nabla(fA_0)=0,\\
\label{fG2}
&\nabla\cdot \Big(\frac{f^2A_0^2}{\mu}\nabla S_0\Big)=0. 
\end{empheq}
This includes the special case in air where the reference wave solution $E_z^0$ in source-free region follows 
\begin{empheq}[left=\empheqlbrace]{align}
\label{mainsG}
\nabla^2 A_0-A_0\nabla S_0\cdot \nabla S_0+ k_0^2 A_0=0,\\
\label{mainG2}
\nabla\cdot \big({A_0^2}\nabla S_0\big)=0. 
\end{empheq}
We then solve the material profiles in Eqs.~\eqref{fG} and \eqref{fG2} as
\begin{empheq}[left=\empheqlbrace]{align}
\label{mu2}
 \mu(\mathbf x)&=f^2, \\
  \label{eps2}
 \epsilon(\mathbf x)&=\frac{1}{k_0^2f^2}\Big(k_0^2-\frac{\nabla^2 f}{f}+\frac{2\nabla f\cdot\nabla f}{f^2}\Big),
\end{empheq}
by making use of Eqs.~\eqref{amp} and \eqref{mainsG}. Here by definition $f$ is positive. Note that this solution is obtained under the condition that the source sits in the far field region of $f(x, y)\sim1$. If the source is put in region $f(x, y)\neq1$, the scattering-free effect of the designed material will no longer hold. The wave amplitude $A(x, y)$ in the whole domain can be designed in an arbitrary manner~\footnote{Due to our condition that the current source sits in trivial region, we cannot control the wave amplitude in the source region. } in the central region when $f\neq 1$. The physical constraint on $f(x,y)$ is that it should smoothly vary within space in order to give a finite range for material profiles in Eqs.~\eqref{mu2} and ~\eqref{eps2}. The function $f$ can vary with a sharper resolution than the wavelength, although resulting in a larger value range for the requisite material profile $\epsilon(x,y)$. We shall give an example for this method to control the wave amplitude in Sec.~\ref{Sec3} where the source is chosen as a point source in 2D space.

\subsection{Controlling the phase}\label{method2}
We can also control the phase function of the wave. If a given phase distribution $\phi(x, y)$ is desired, we can use Eq.~\eqref{phase} to find the requisite material parameters. We solve for wave amplitude and material parameters, starting from Eq.~\eqref{mains2} and write it compactly by defining $F:=\ln\zeta:=\ln{A^2}/{\mu}$. One thus has
\begin{eqnarray}
 \nabla F\cdot \nabla \phi&=&-\nabla \cdot \nabla \phi, 
 \end{eqnarray}
 Then we may write 
 \begin{eqnarray}
 \label{integrability}\nabla F&=&-\frac{\nabla^2\phi}{\vert\nabla \phi\vert^2}\nabla \phi+K(x, y)\mathbf{p}\\
 \label{simple}&\approx& -\frac{\nabla^2\phi}{\vert\nabla \phi\vert^2}\nabla \phi\quad(\mathbf{p}\cdot\nabla \phi=0). 
\end{eqnarray}
Above $\mathbf{p}$ is a unit perpendicular vector to $\nabla \phi$ and $K(x, y)$ is an unknown factor in front of $\mathbf{p}$~\footnote{Note that the variable $K$ as the factor before the perpendicular vector $\mathbf{p}$ is actually \emph{not} arbitrary as it might appear at the first sight. The reason not being so is two-fold: 1) the requirement that material parameters reduce to unity prevents us from exploiting the freedom; 2) moreover, $K(x, y)$ is restrained in order to compliment the former term on the right side of Eq.~\eqref{integrability} into a gradient. Therefore it \emph{cannot} be arbitrary function of $x$ and $y$. Thus Eq.~\eqref{integrability} is an integrability problem to solve. Once $F(A^2, \mu)$ is known, one can design the wave amplitude $A$ and the permeability becomes known. }. Though generally the partial differential equation for $F$ could be solvable by numerics with an appropriately chosen boundary condition, we choose to approximate the solution as Eq.~\eqref{simple} because it lends a simpler closed form in material parameters. 

For simplicity, in non-magnetic case ($\mu=1$) we can approximate the permittivity in an analytic form by substituting Eq.~\eqref{simple} into Eq.~\eqref{mains1}, and then one obtains
\begin{eqnarray}
\label{epsf}
\epsilon&\approx&\frac{1}{k_0^2}\Bigg[\vert\nabla \phi\vert^2
+\Big(\frac{\nabla^2\phi}{2\vert\nabla \phi\vert}\Big)^2+\nonumber\\
&&\frac{\nabla \phi}{2\vert\nabla \phi\vert^4}\cdot\bigg(\vert\nabla \phi\vert^2\nabla\nabla^2\phi-2\nabla^2\phi(\nabla \phi\cdot\nabla)\nabla \phi\bigg)\Bigg].
\end{eqnarray}
The first leading term $\vert\nabla \phi\vert^2/k_0^2$ in Eq.~\eqref{epsf} is essential from the geometrical-optical eikonal equation. The rest terms represent the high order correction to give a more accurate design to account for wave optics than for geometrical optics. Note that generally $\nabla \phi$ no longer equals $k_0$ as the module of wave vector varies through the inhomogeneous medium. One can also predetermine magnetic material $\mu\neq 1$ and then obtain a different permittivity $\epsilon(x, y)$, in which the same phase as nonmagnetic material is achieved. We will give two examples of wave front conversion design in Subsec.~\ref{converter1sub}.

\section{\label{Sec3}Invisible material for a point source}

\begin{figure*}[h]

\begin{center}
\includegraphics[width=0.47\textwidth]{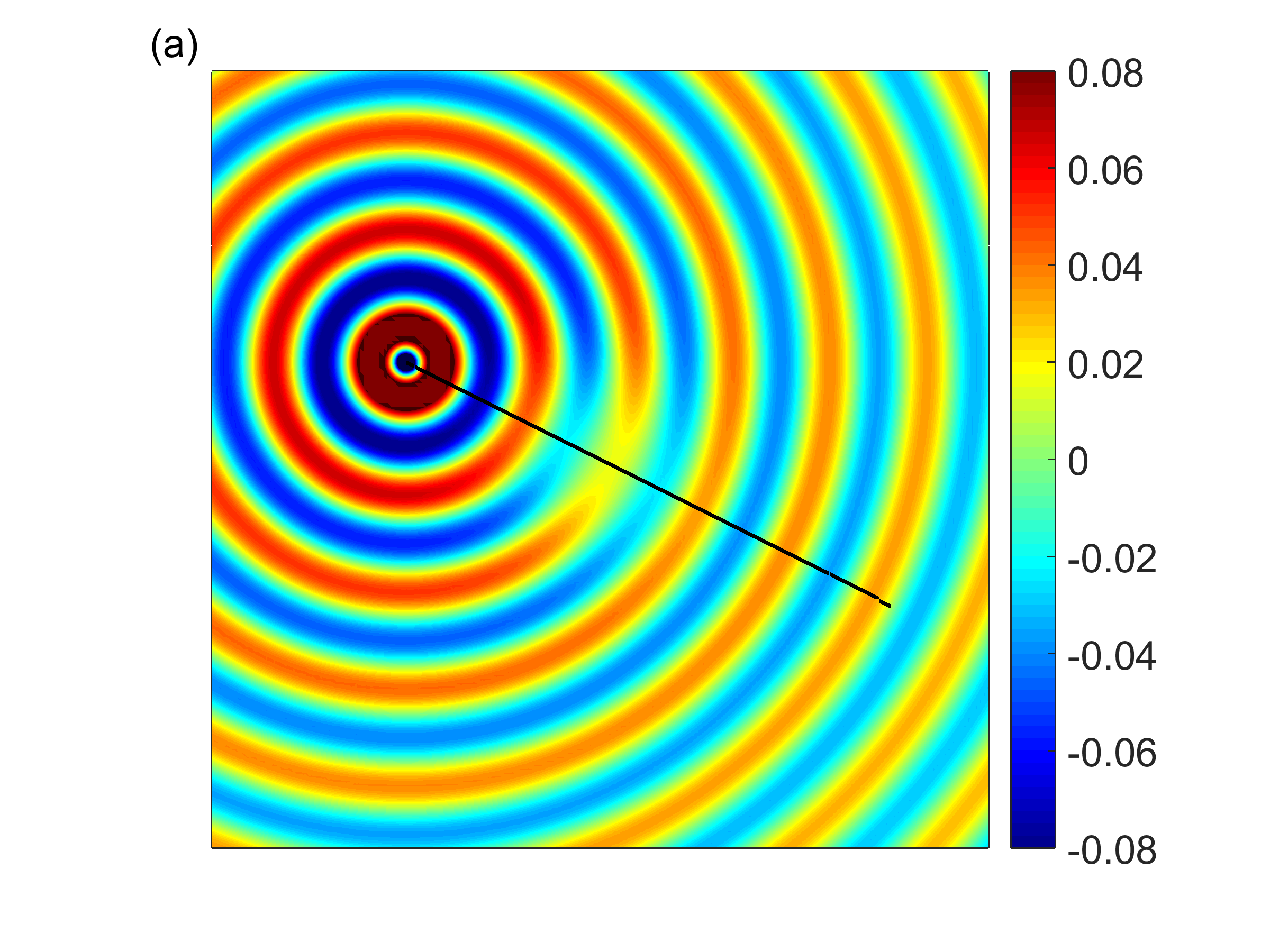}
\includegraphics[width=0.47\textwidth]{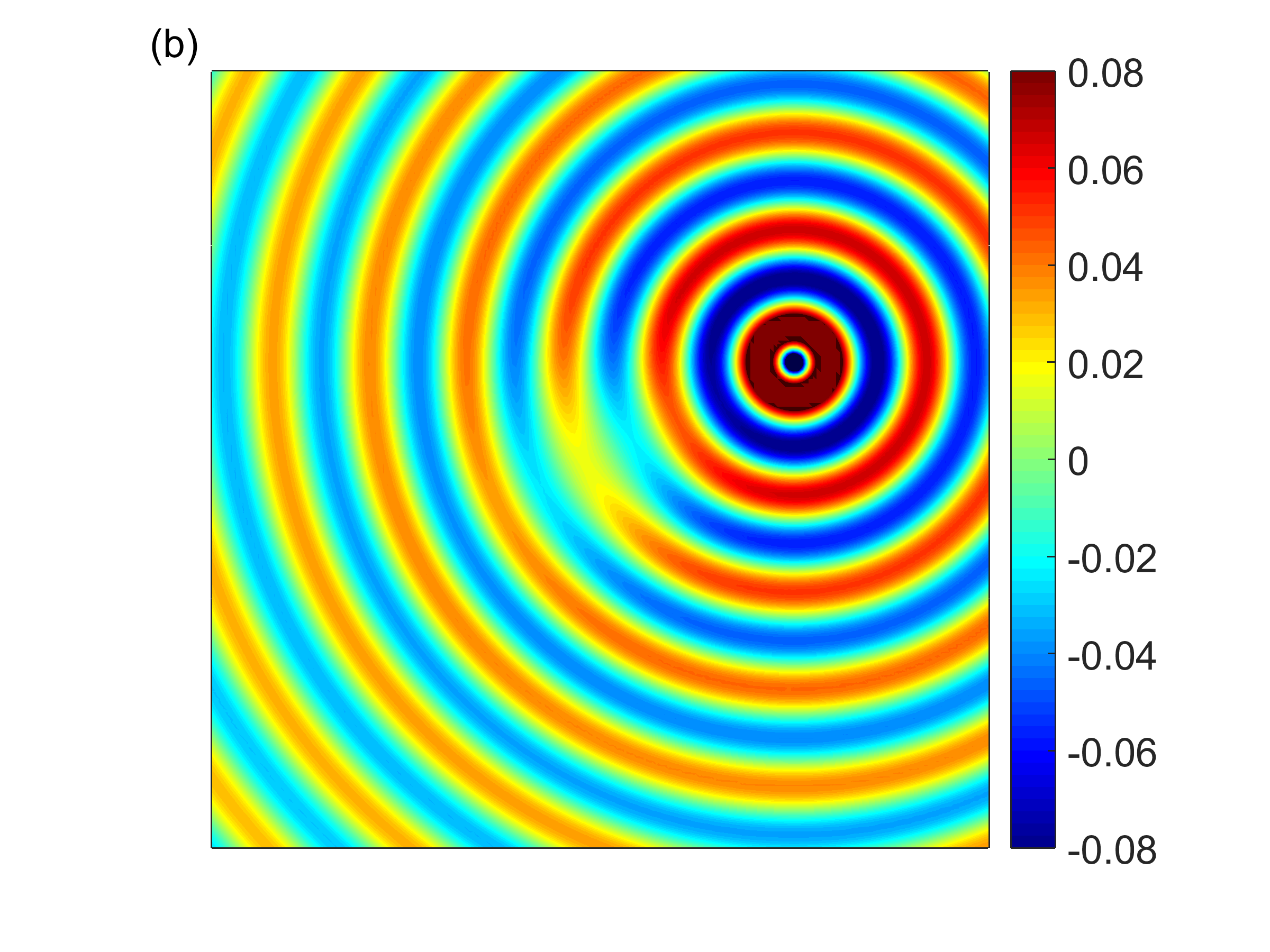}\\
\includegraphics[width=0.47\textwidth]{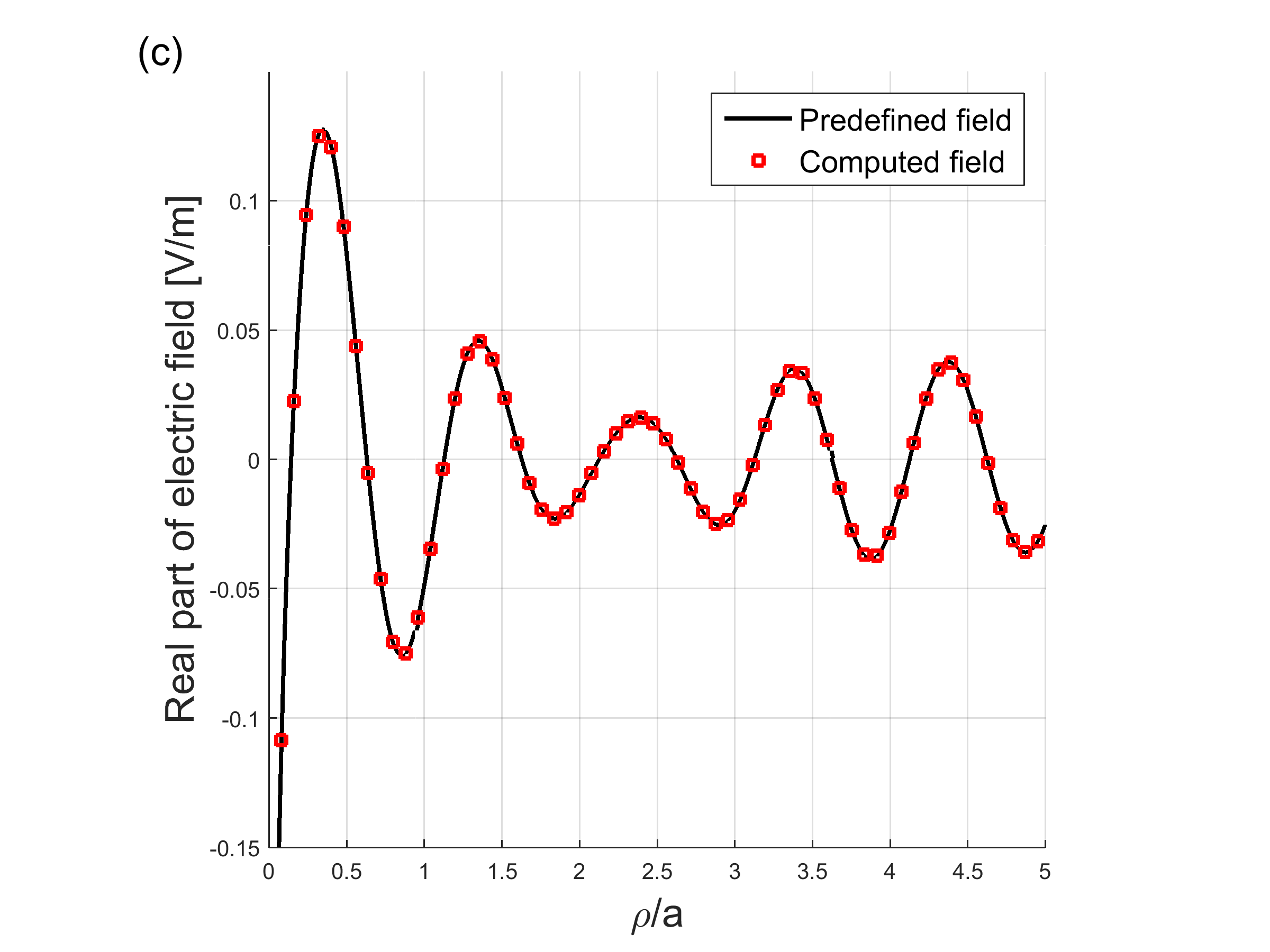}
\includegraphics[width=0.47\textwidth]{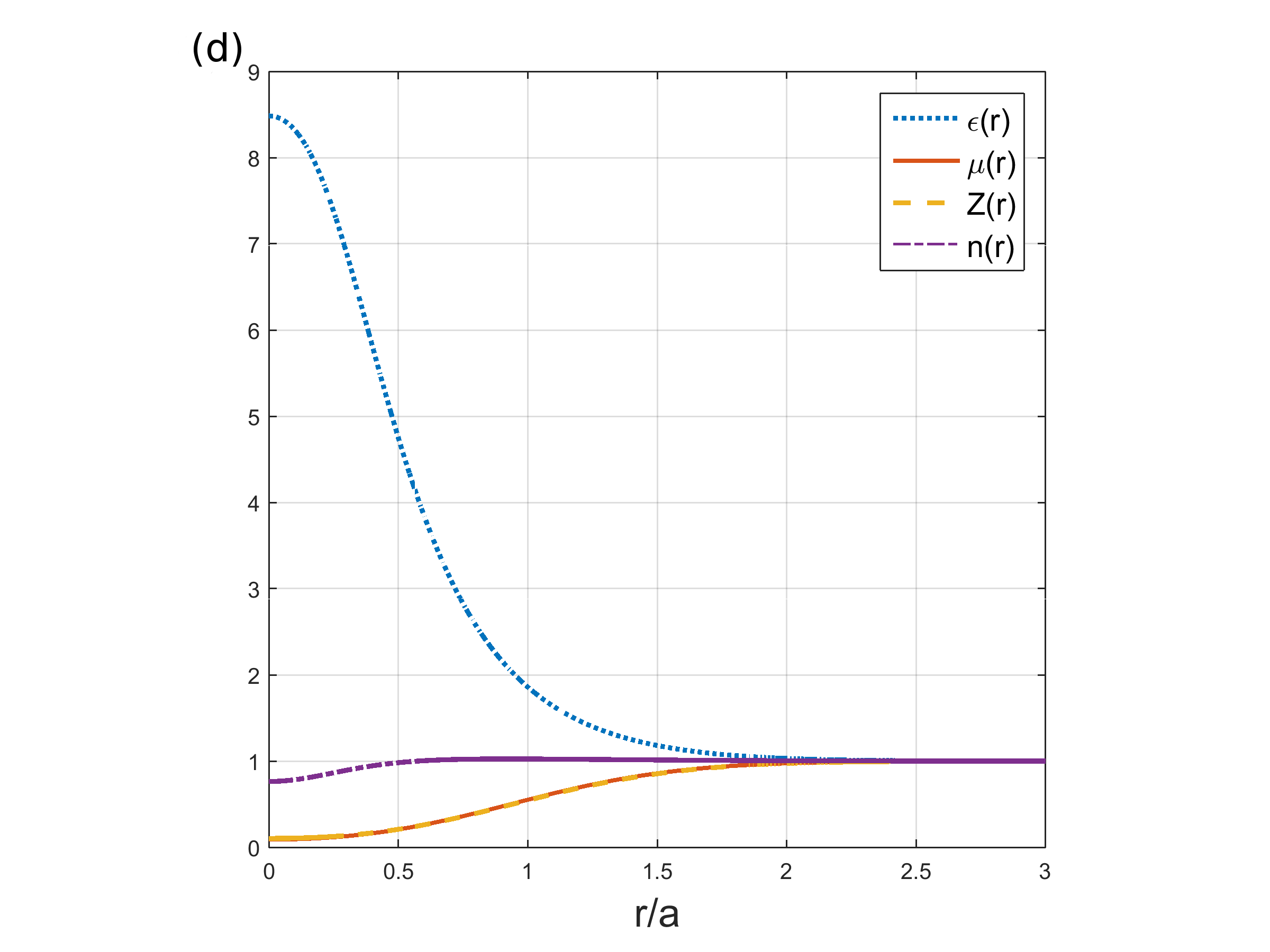}

\caption { \label{fig:Fig2} Invisible material for two cylindrical waves from points $(b, c)$ and $(-b, c)$. (a-b) computed wave solutions when light emits from (a) $(b, c)$ and (b) $(-b, c)$ respectively; (c) comparison of electric field values between predefinition and computation along the black line segment in (a); (d) designed permittivity $\epsilon(r)$, designed permeability $\mu(r)$, impedance parameter $Z:=\mu/\epsilon$ and refractive index ($n:=\sqrt{\mu\epsilon}$). Parameters: $a=1{\rm m}, \alpha=0.7, \lambda=a, b=-2, c=1$. Produced from \texttt{COMSOL} 5.2 simulation. 
 }

\end{center}
\end{figure*}

Apart from planar propagation~\cite{Ben2016b}, a simple incident wave in 2D case is the cylindrical wave from a point source (a line of source in 3D space). As an example to apply the method in Sec.~\ref{method1}, in the same material Eqs.~\eqref{mu2} and \eqref{eps2}, the amplitude of a cylindrical wave is also controllable. In this case the reference wave $E_z^0$ in Eq.~\eqref{refE0} becomes the Green function for a point source, which is written as 
\begin{eqnarray}
\label{Ez0}
E_z^0(x, y)&=&\frac{i}{4}(J_0(k_0\rho)-iY_0(k_0\rho))\\&=&\frac{1}{4}\sqrt{J_0(k_0\rho)^2+Y_0(k_0\rho)^2}\cdot\exp (-i\phi_i)\nonumber\\\Big(\rho:&=&\sqrt{(x+b)^2+y^2}\Big), 
\end{eqnarray}
where $J_0(\cdot), Y_0(\cdot)$ are respectively, the first and the second kinds of Bessel functions of zeroth order and the cylindrical phase $\phi_{\rm i}$ is defined as ${\arctan}[Y_0(k_0\rho),-J_0(k_0\rho)]$. The controlled amplitude is thus
\begin{eqnarray}
A=fE_z^0. 
\end{eqnarray}
As an example we indicate $f=1-\alpha e^{(x^2+y^2)/a^2}$ where $\alpha, a$ are parameters. The wave solution and the scattered wave from a point source in such a reflectionless medium is plotted in Fig.~\ref{fig:Fig2}(a-b), respectively. It is noted that our designed material Eqs.~\eqref{mu2} and \eqref{eps2} plotted in Fig.~\ref{fig:Fig2}(d) works for any point source positioned in free space region. The excellent comparison between simulation and predefined wave field value along a line segment in Fig.~\ref{fig:Fig2}(a) is demonstrated in Fig.~\ref{fig:Fig2}(c).

Note that in fact, our designed material in Eqs.\eqref{mu2} and \eqref{eps2} works for any current source $j(x, y; \omega)$ sitting anywhere but the non-trivial region where factor $f$ strongly deviates from unity, {\emph i.e.} the designed medium independent on the source modulates the field from any source, which causes no scattering throughout the space. Hence this is actually another class of invisible media made of positive isotropic material similar to ~\cite{Tyc2014b} but possesses \emph{zero} phase delay, although no room to hide any object in this case. Generally, the rotational symmetry of $f(x, y)$ is not a compulsory condition and a more general profile still works as an invisible material, as long as the material profiles are continuously second-order derivable.

For summary of Sec.~\ref{Sec3}, we give an examples of reflection-less wave from a point source, which enters the medium from one side and recovers to its initial wave shape at the other side. 

\section{Phase conversion material}\label{converter}

In this Sec., we will design material profiles which convert an incident wave of one phase into another. The examples are given in Subsec.~\ref{converter1sub}. The first converts a cylindrical wave to a planar one, and the second to another shifted cylindrical wave, as if from an illusional point source.  In Subsec.~\ref{Luneburg} we also validate our method to control the phase, reproducing L\"uneburg lens as a special case.

\subsection{Wave front conversion - from a cylindrical wave to other phases}\label{converter1sub}

We apply the general formula Eq.~\eqref{epsf} derived in Subsec.~\ref{method2} to design a conversion material, which transfers a cylindrical wave from a point gradually to another wave of phase. The phase $\phi(x, y)$ of the wave in such a medium
can be written as
\begin{eqnarray}
\label{Sconv}
\phi(x, y)=[1-\tau(x)]\phi_{\rm i}(x, y)+ \tau(x)\phi_o(x, y). 
\end{eqnarray}
Here $\phi_{\rm i}$ is the incoming wave phase, $\phi_{\rm o}$ the output wave phase and $\tau(x):=[{1+\tanh(\beta x)}]/{2}$ is a transition function varying from 0 to 1 as $x$ increases~\footnote{We choose only x dependence for simplicity. }. In $\tau(x)$ free parameter $\beta$ can be tuned to control how quickly the phase varies from cylindrical (the incident wave phase $\phi_{\rm i}$ is cylindrical according to Eq.~\eqref{Ez0}) to planar. Then we obtain immediately the permittivity profile from Eq.~\eqref{epsf}. 

\begin{figure*}[htbp]

\begin{center}

\includegraphics[width=0.495\textwidth]{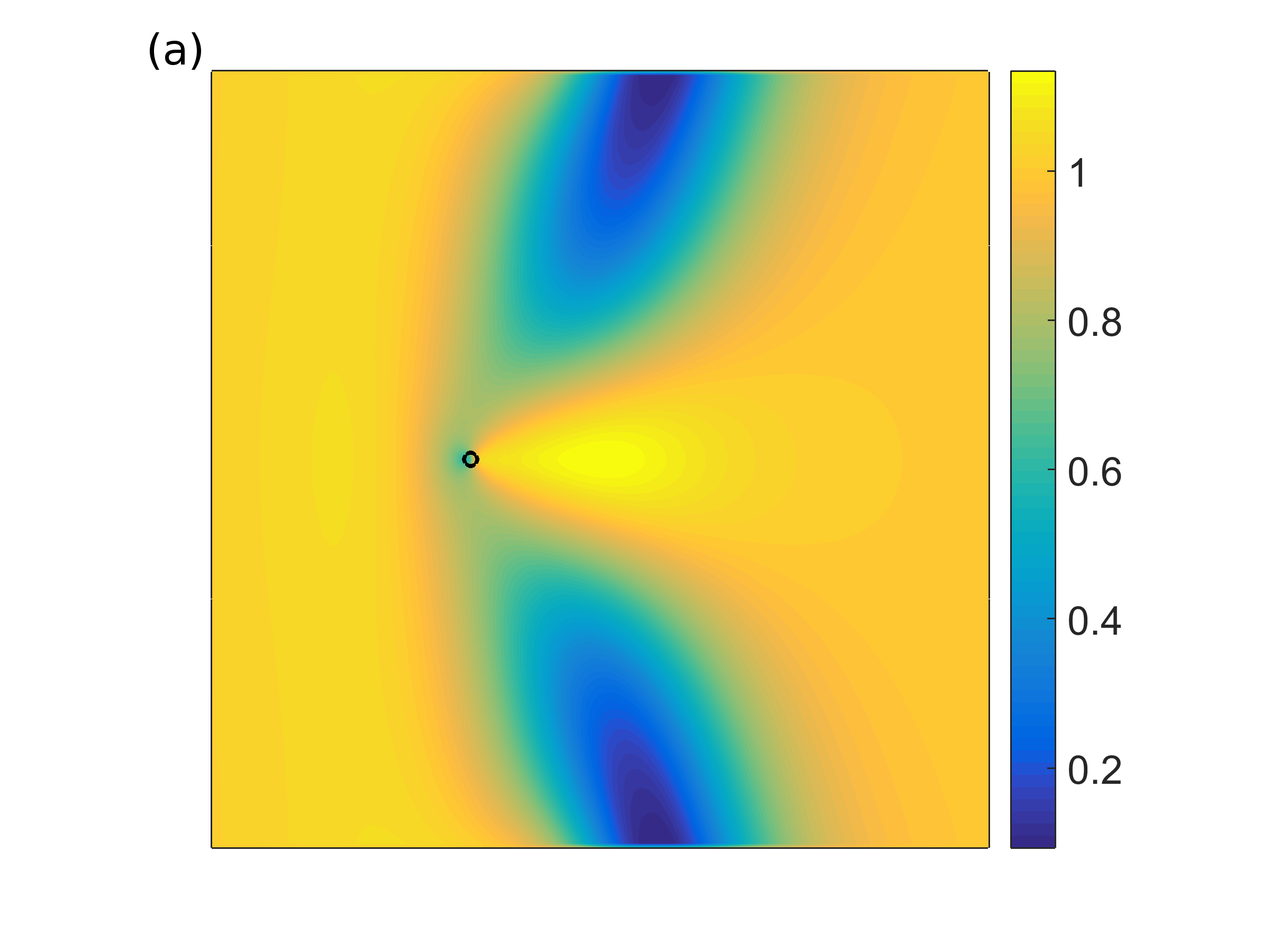}
\includegraphics[width=0.495\textwidth]{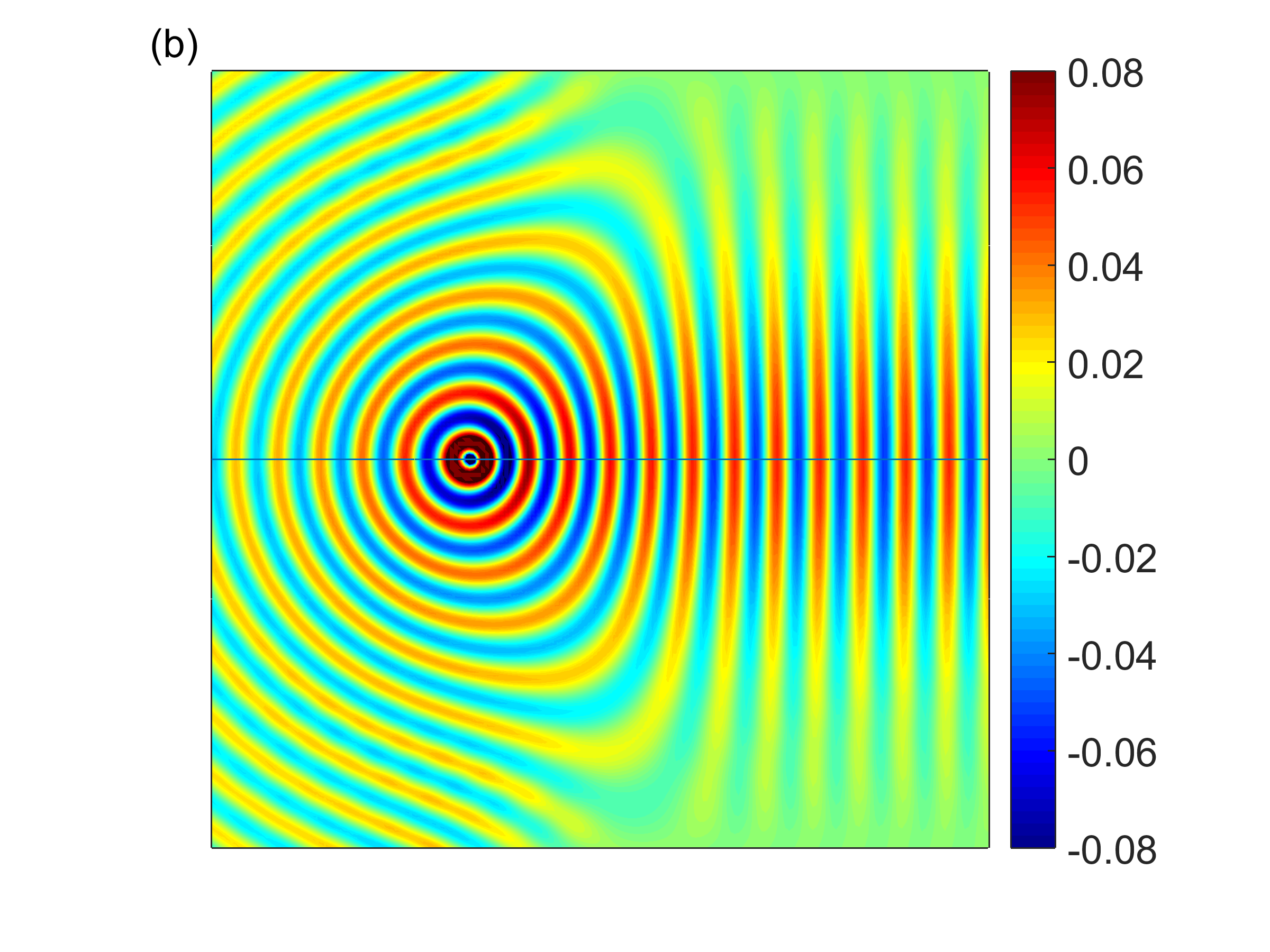}\\
\includegraphics[width=0.495\textwidth]{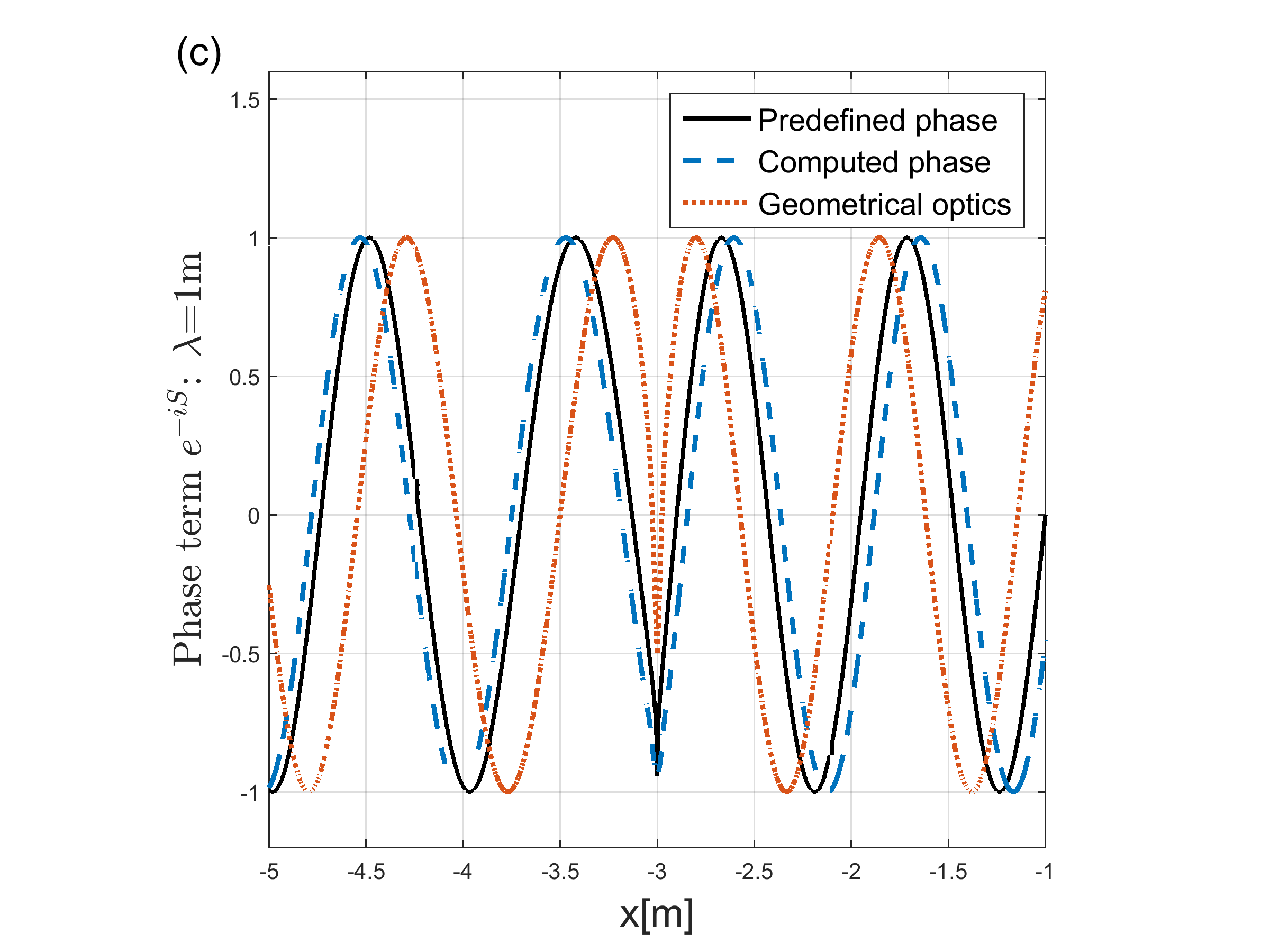}
\includegraphics[width=0.495\textwidth]{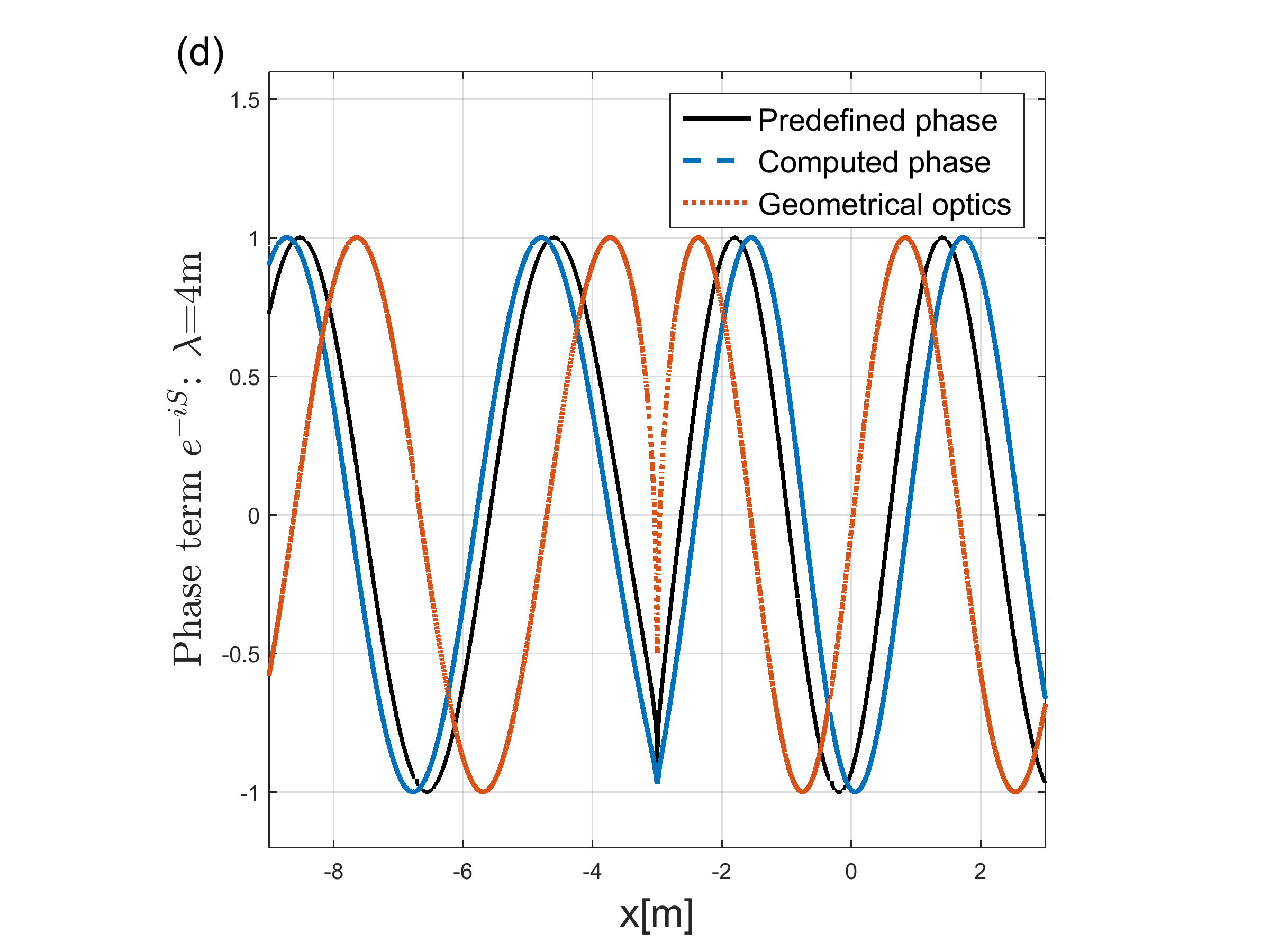}\\
\includegraphics[width=0.495\textwidth]{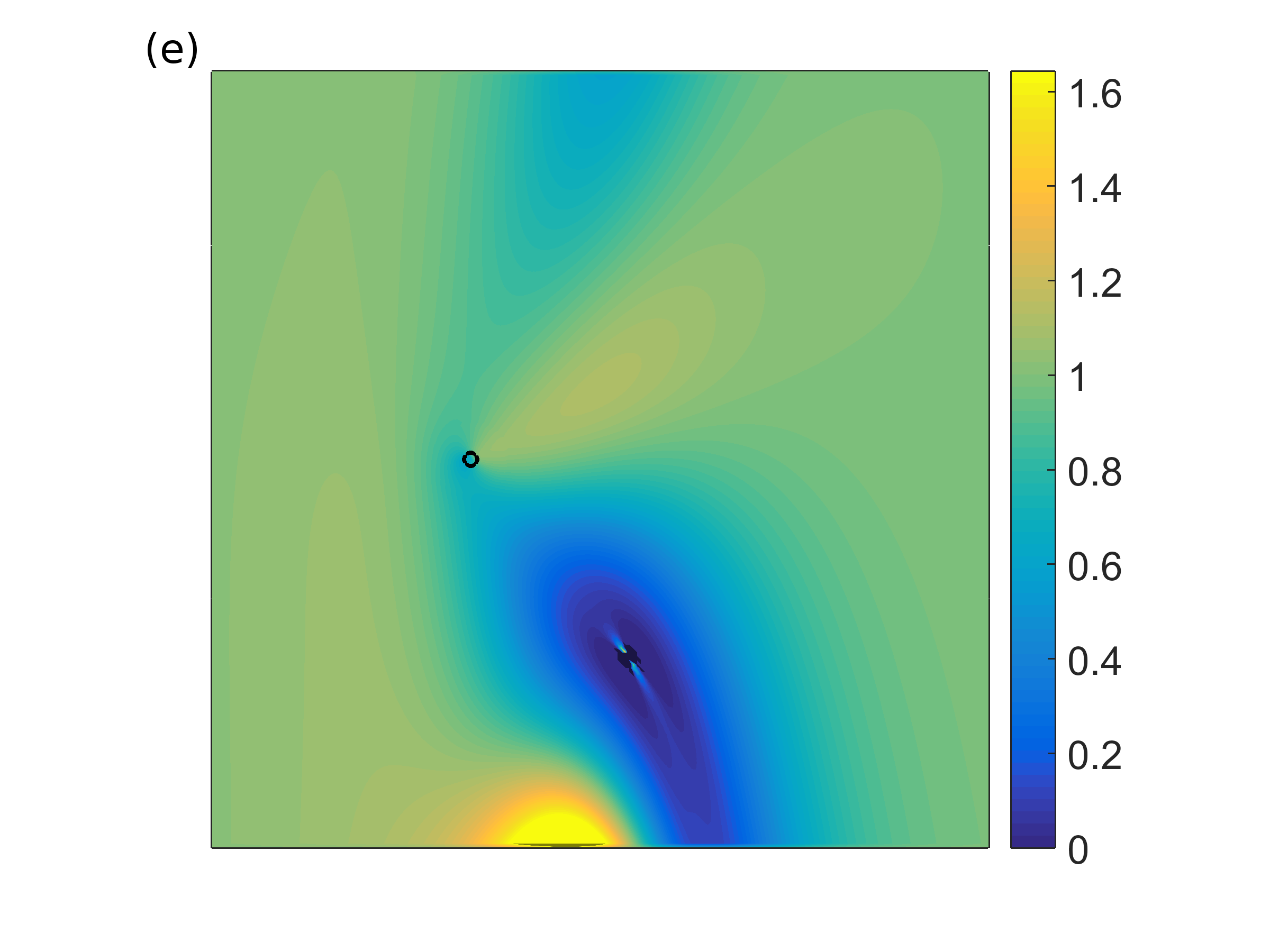}
\includegraphics[width=0.495\textwidth]{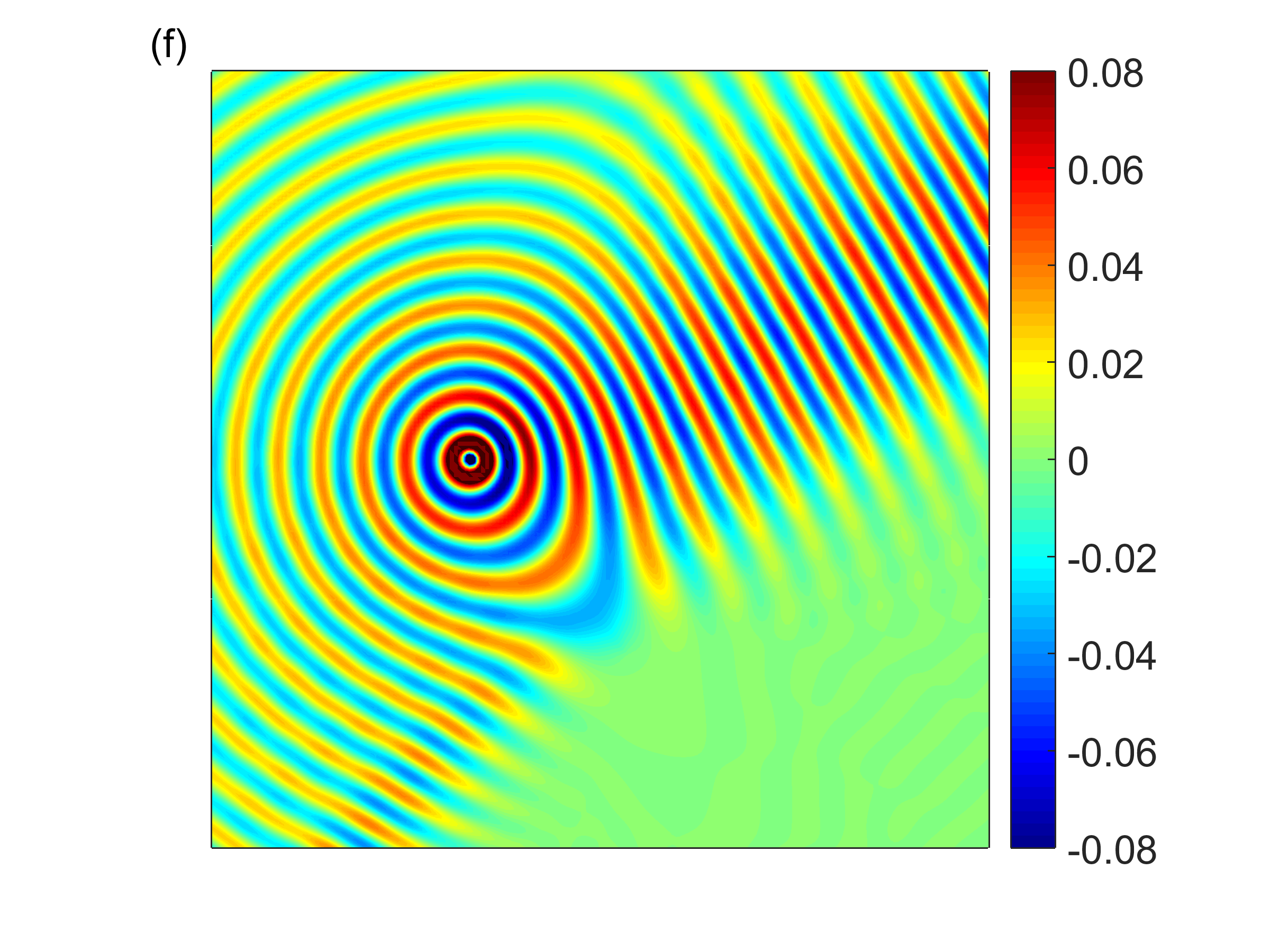}

\caption { \label{fig:Fig3} Phase conversion design for two transmittance angles: for  (a, e) designed profile of permittivity; (b, f) electric field plot. (a, e): the point sources in black circles. (c-d) Phase plots for (a) along the black line therein, compared with predefined phase Eq.~\eqref{Sconv} and the computed wave phase from a design permittivity from geometrical optics $\epsilon=\vert\nabla \phi\vert^2/{k_0^2}$. Parameters: $\beta=0.35, l=2$; (a-b) $\theta=0$, (c-d) $\theta=\pi/6$.  Produced from \texttt{COMSOL} 5.2 simulation. }

\end{center}
\end{figure*}

First we choose the output beam to be a wave of planar phase~\footnote{Reciprocality of wave equation implies that one could also choose the output to be a cylindrical wave. However, this will give a wave picture of inevitable scattering (not shown in the paper). }
\begin{eqnarray}
\phi_{\rm o}:=k_0[(x+b)\cos\theta+y\sin\theta], 
\end{eqnarray}
in which $\theta$ is the angle between incident wave vector and x axis. Numerical wave simulation in Fig.~\ref{fig:Fig3} shows that one can design the conversion medium not only for normal incidence angle depicted in Fig.~\ref{fig:Fig3}(a-b), also for an oblique one as in Fig.~\ref{fig:Fig3}(e-f). For the case  of normal incidence, we compare the computed phase $e^{-i\phi}$ along x axis (see black line in Fig.~\ref{fig:Fig3} (b)) from simulation with the predefined phase in Eq.~\eqref{Sconv}. In Fig.~\ref{fig:Fig3} (c-d) the  phase from our design differs only by a small delay, and the phase from geometrical optical design $\epsilon=\vert\nabla \phi\vert^2/k_0^2$ is more errant from predefined phase. Thus our method based on wave optics yields a more accurate medium than from geometrical optics. 

\begin{figure*}[htbp]

\begin{center}
\includegraphics[width=0.45\textwidth]{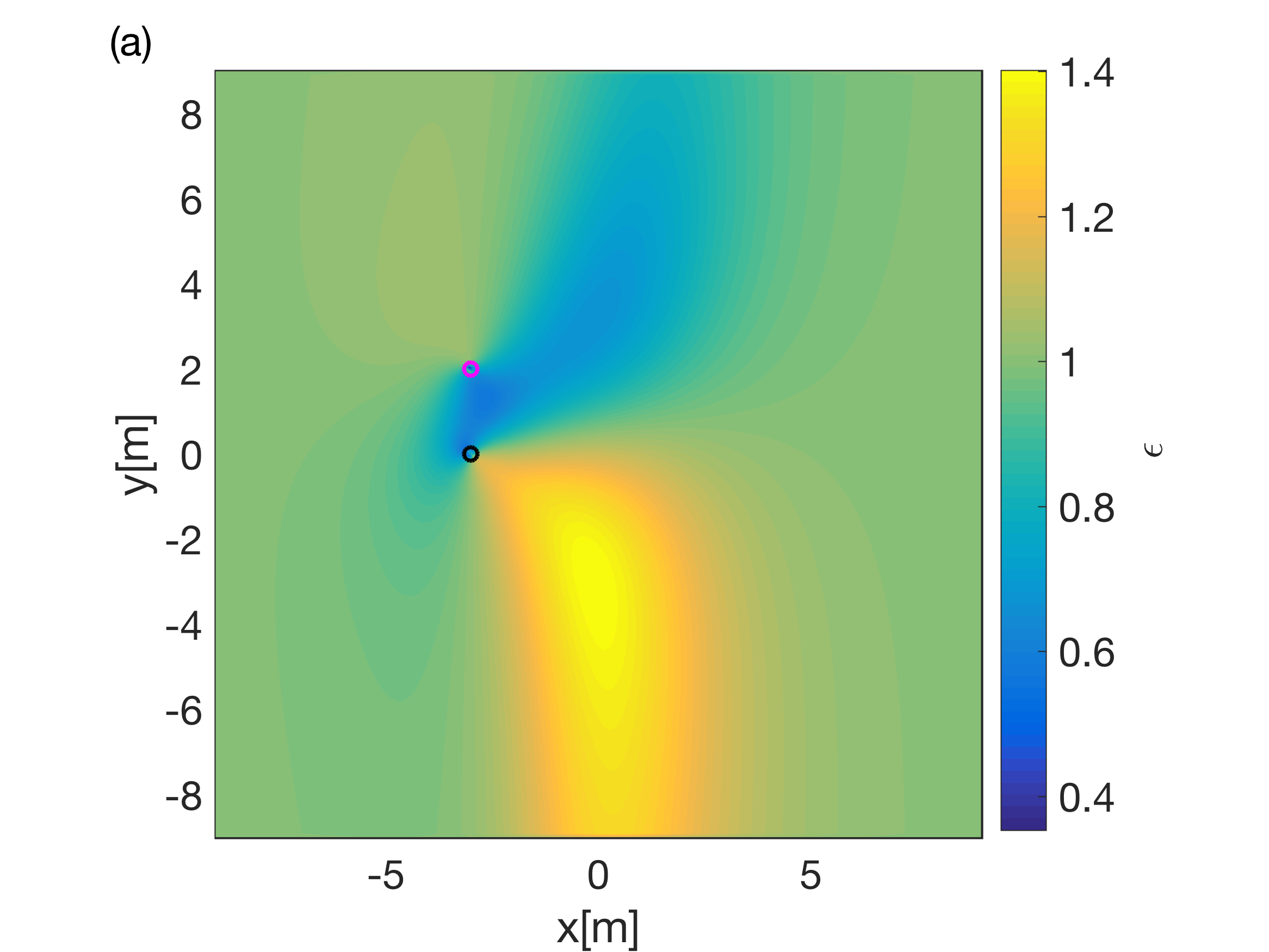}
\includegraphics[width=0.45\textwidth]{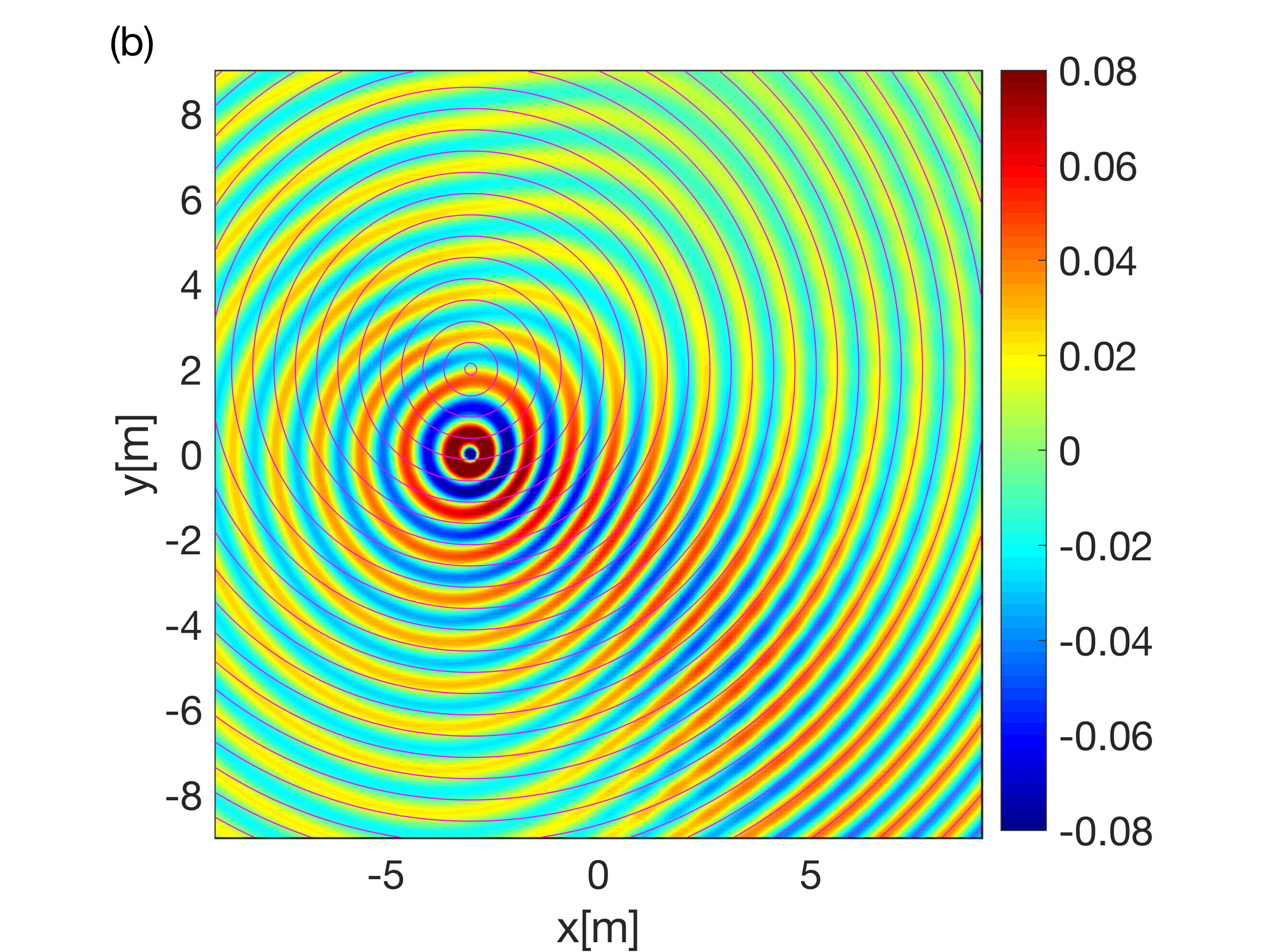}\\
\includegraphics[width=0.45\textwidth]{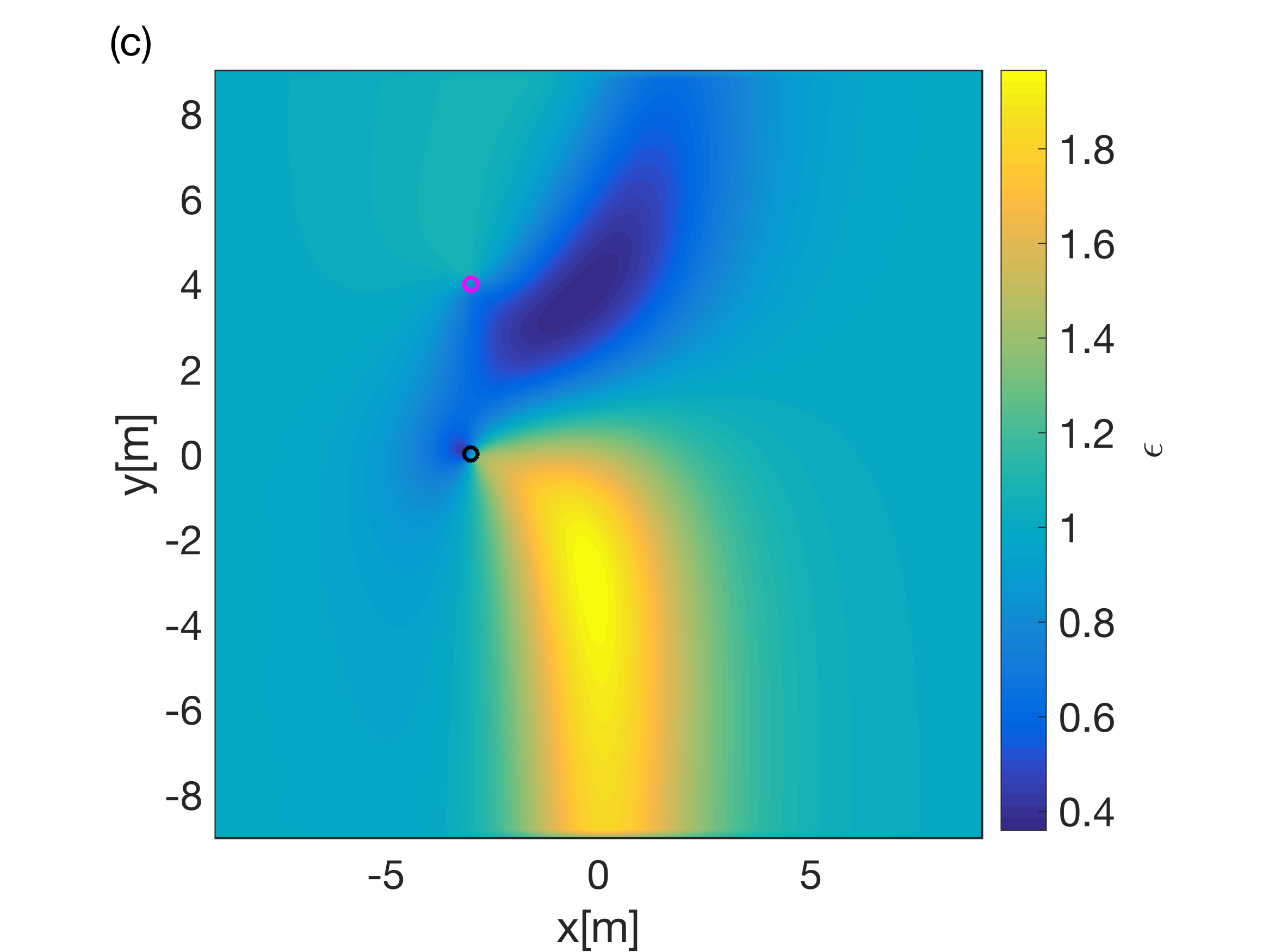}
\includegraphics[width=0.45\textwidth]{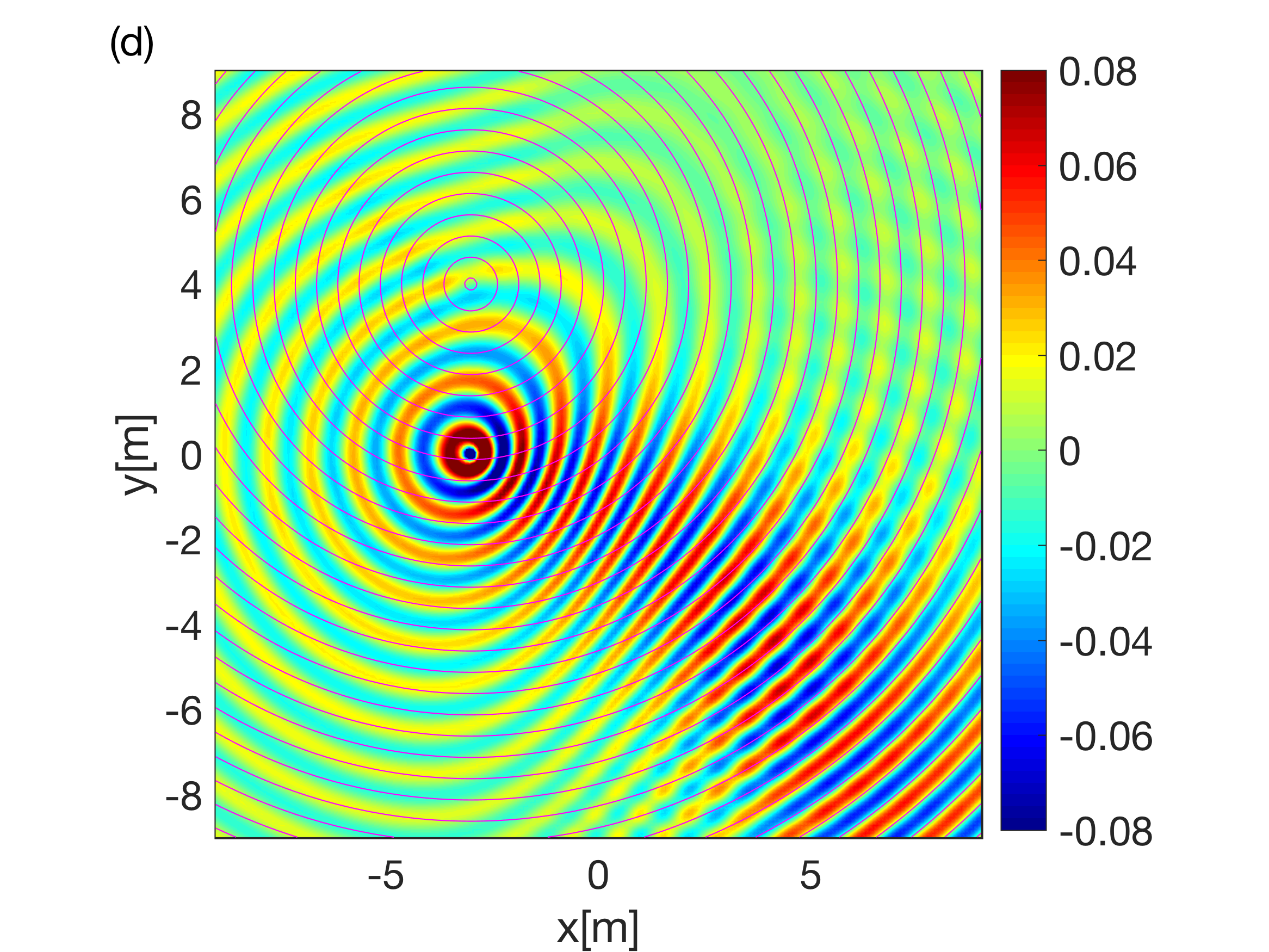}

\caption { \label{fig:Fig5} Beam shifting material design: (a, c) designed permittivity profile map $\epsilon(r)$, the point source in black circle; (b, d) electric field plot in the beam shifter medium, along with the field contours from the illumination point (-b, c) in magenta. Parameters: (a-b) $b=3, c=2$; (c-d) $b=3, c=4$. Produced from \texttt{COMSOL} 5.2 simulation.   }
\end{center}
\end{figure*}

Secondly, we also design a beam shifting medium similar to \cite{Gallina2010} where the output wave phase is chosen to be a laterally-shifted cylindrical phase from $(-b, c)$: 
\begin{eqnarray}
\phi_{\rm o}:={\arctan}[Y_0(k_0\rho'),-J_0(k_0\rho')]\\\nonumber (\rho':=\sqrt{(x+b)^2+(y-c)^2}). 
\end{eqnarray}
From simulation, the designed profiles of permittivity in Fig.~\ref{fig:Fig5}(a, c) transfers the wave from the point source (in black) to that from the shifted one (virtual, in magenta). The phase of the output beam from the point source converts to that of a shifted beam, which are indicated by magenta contours in Fig.~\ref{fig:Fig5}(b, d).

\subsection{An example to illustrate the validation of phase conversion materials}\label{Luneburg}
\begin{figure*}[htbp]

\begin{center}
\includegraphics[width=0.5\textwidth]{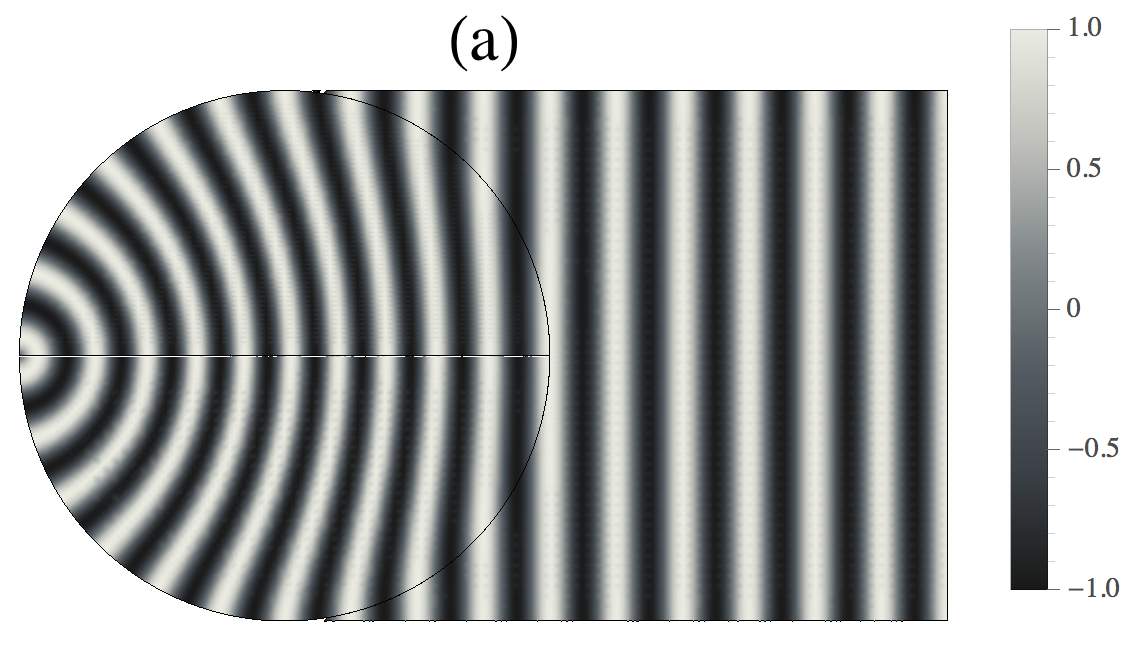}\\
\includegraphics[width=0.5\textwidth]{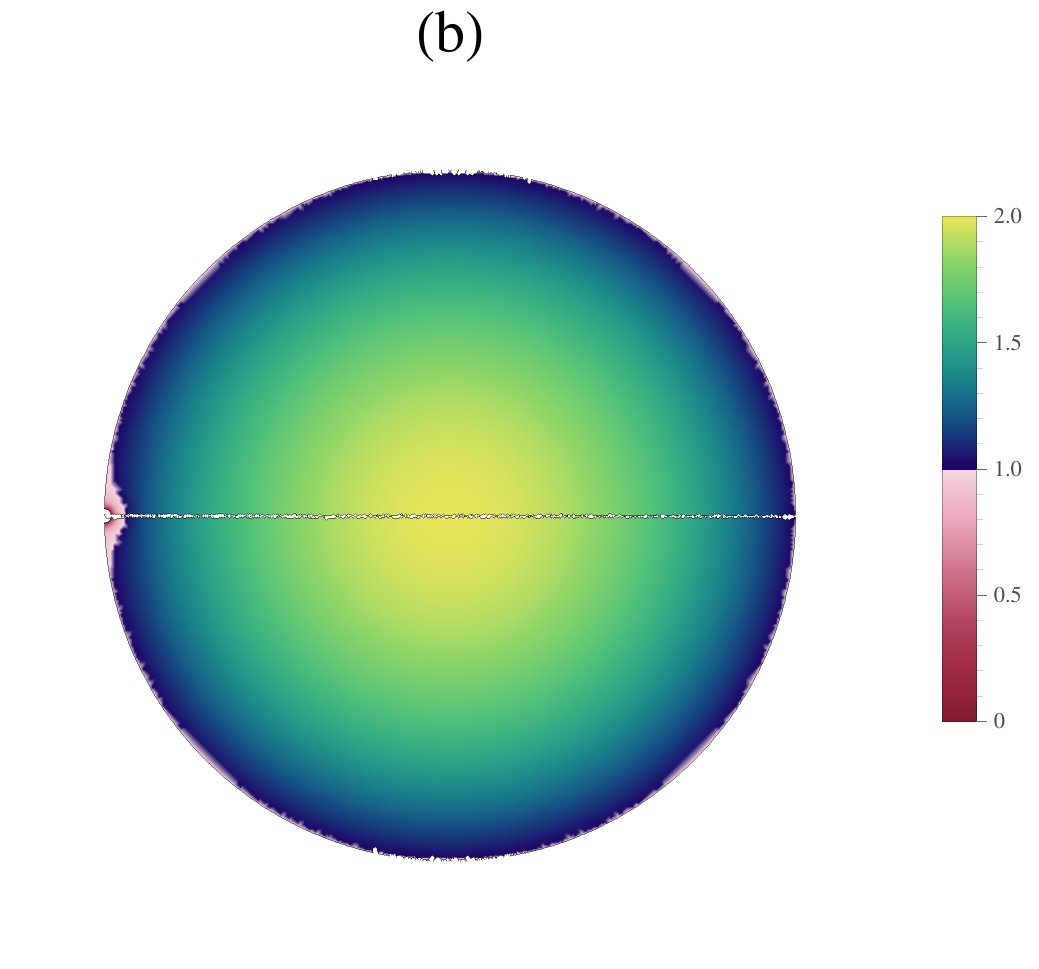}\\
\includegraphics[width=0.43\textwidth]{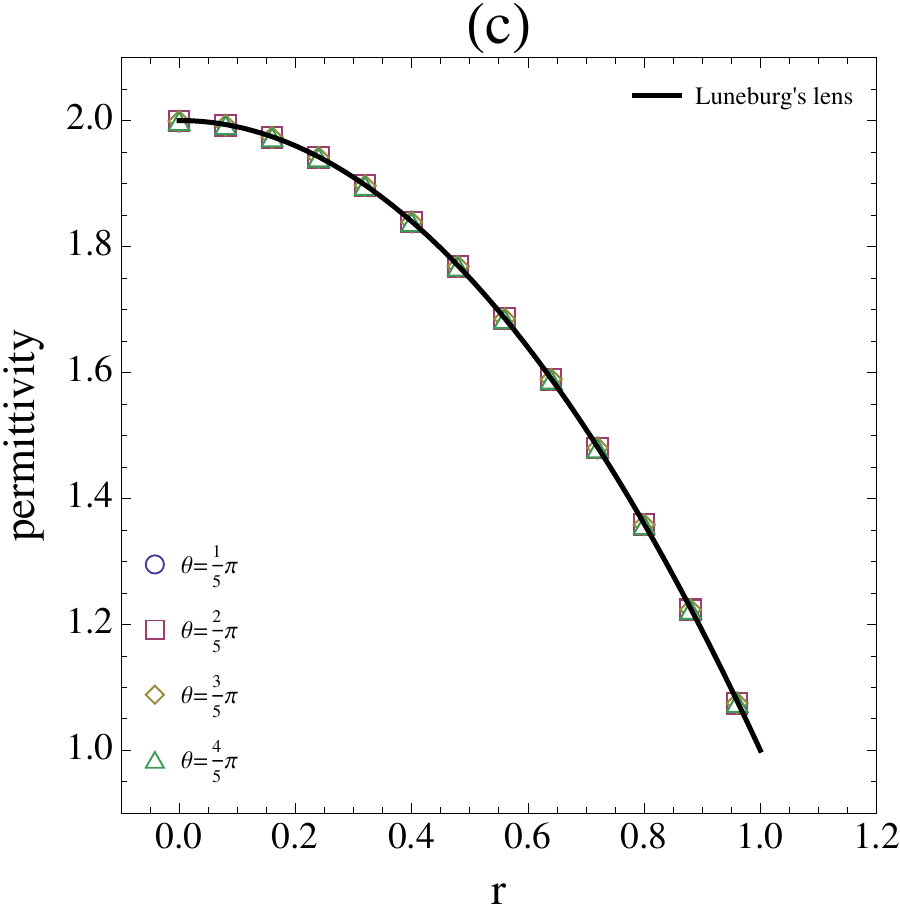}

\caption { \label{fig:Fig4} Recovered permittivity profile of L\"uneburg lens: (a) analytical phase distribution; (b) permittivity profile map; (c) permittivity function $\epsilon(r)$ along different azimuthal angle $\theta$. Computed from \texttt{Mathematica} code. }

\end{center}
\end{figure*}

As the L\"uneburg lens focuses a planar wave into a single point at its edge, we naturally wonder if this is recoverable via our design formula~Eq.~\eqref{epsf}. In the case of planar-wave illumination, we will retrieve the analytic phase distribution (see Fig.~\ref{fig:Fig4}(a)) from ray-tracing~\cite{Leonhardt2010book}, and obtain an \emph{almost} identical profile of permittivity as L\"uneburg lens, depicted in Fig.~\ref{fig:Fig4}(b-c). Further detail is included in Appen.~\ref{app2}. We are therefore assured that this method is reliable.


\subsection{Metamaterial implementation}
In order to verify feasibility of our design methodology, we briefly discuss a possible metamaterial implementation here. Our previous implementation and simulation already verify our design methodology~\cite{Ben2016b}(Sec. IIC therein) and thus we shall not repeat in detail. We propose to use two kinds of materials (SiC silicon carbide  and KBr potassium bromide with permittivity $\epsilon_{\rm SiC}=0.0009 +i0.0815$ and $\epsilon_{\rm KBr}=2.3280$ at $10.32\mu m$), which are hosted in air to arrange as small-rod arrays. One can compute which requires the rod-sizes from homogenisation theory (Maxwell-Garnett formula).  In engineering realisation, the material loss could play a role and weaken the wave amplitude than predicted~\cite{Ben2016b}. Note that our proof-of-concept design is \emph{de facto} based on lossless material parameters, which can be taken into account to give rise to an more accurate design scheme.

\section{Deviation from exact geometrical optics}\label{Ego}
A natural question to ask is whether our method aligns with geometrical optics~\cite{Philbin2014}. For TE wave, we find that geometrical optics stands when
\begin{equation}\label{GO}
\nabla\cdot \frac{\nabla A}{\mu}=0, 
\end{equation}
such that Eq.~\eqref{mains1} recovers the eikonal equation~\cite{Philbin2014}
\begin{eqnarray}
\label{eikon}
-\nabla \phi\cdot\nabla \phi+n^2 k_0^2 =0, 
\end{eqnarray}
if one defines effective refractive index $n:=\sqrt{\epsilon\mu}$. 
Therefore it is easy to find that both our examples in Fig.~\ref{fig:Fig2} are deviant from the condition of exact geometrical optics~\footnote{see Eq.~(29.4) of \cite{Leonhardt2010book}}. Moreover for TE wave, Eq.~\eqref{eikon} \emph{will} stand for all frequency, if a different wave equation 
\begin{eqnarray}
\label{Ez2}
\nabla^2 E_z+\epsilon\mu k_0^2 E_z-\frac{1}{\mu}\nabla \mu\cdot\nabla E_z-\mu\frac{E_z}{\vert E_z\vert}\nabla\cdot\frac{\nabla\vert E_z\vert}{\mu}=0
\end{eqnarray}
can be postulated, if we are able to allow some special material response which contributes to the last term above~(\emph{cf}. Sec. 5 of ~\cite{Philbin2014}).

\section{Conclusion}\label{conc}
In summary, this paper extends our method~\cite{Ben2016b} to design for scattering-free wave amplitude or phase in isotropic, heterogeneous medium and focus on manipulating cylindrical wave in 2D case. This elegant analytic method inversely designs the material parameters in a different  route compared with transformation optics~\cite{Leonhardt2006a, Pendry2006}. We give two types of examples to demonstrate our design scenario: an invisible material and a wave front conversion medium both for a point source. This method generally works for monochromatic frequency and is highly narrow-band limited. Our design method and and simulation verification are given for 2D TE wave for isotropic medium, which can be extended to the TM wave, and even 3D case~\footnote{See \cite{Potvin2015} for a mathematical approximation in 3D case. }. Extension to anisotropic materials is also possible. Moreover, our method could camouflage a certain amount of scattering wave, although the inverse problem for a scattering wave should be far more complicated and challenging. We hope this method can shed some light on how to design graded-index materials in order to keep or change the phase of a cylindrical wave.

\begin{acknowledgments}

We are supported by the Engineering and Physical Sciences Research Council (EPSRC) of UK [grant number EP/I034548/1]. L Y benefits from LMS-EPSRC Durham Symposium 104, LMS Caring Supplementary Grant [Ref. CC-15/16-09] and Antennas group travel grant [EECSRC2/4a], and wants to thank Patrick Bradley, Flynn Castle and Fang Yanlong for inspiring conversations to improve this manuscript. This article is supported by the EECS Open Access group and Queen Mary Library ($\#67$).

\end{acknowledgments}

\appendix

\section{Derivation to recover Luneburg lens}\label{app2}

In Subsection ~\ref{Luneburg}, we recover the permittivity profile of L\"uneburg lens by taking the phase function $S(x, y)$ for planar incidence wave illuminated from right side of the lens. Here is the detailed derivation to retrieve the phase information. While one can generally relies on numerical method of solving Hamilton's equations to retrieve the phase information, we choose to analytically solve the phase distribution for two dimension as follows, which shall be easily extended to a three dimensional case.  
Firstly, L\"uneburg lens profile in dimensionless unit is written as 
\begin{equation}
\epsilon(x,y)=n^2(x,y)=2-({x^2+y^2}).
\end{equation}
And we illuminate a planar wave from straight line $x=b$ and focus it into its focus at $(-1,0)$. We shall use complex coordinates $z=x+iy$ to simplify our derivation. What remains is to write explicitly the phase for each point $z(\eta)=u+iv$ inside the unit circle, since the phase outside it is just a planar wave phase. For each point $z(\eta)$, it shall fall on a specific ray penetrating the circle at two intersection points $x_p(u, v)\pm i\sqrt{1-x_p^2(u, v)}$ and $-1$, sequentially. In complex coordinates $z=x+iy$, any tray inside L\"uneburg lens can be written as
\begin{equation}
\label{zxi}
z(\xi')=e^{i\alpha(u, v)}A(u, v)\cos\xi'+iB(u,v)\sin\xi',\;A^2+B^2\equiv 2. 
\end{equation}
And its two joint points with the unit circle are represented as  $z(-{\pi}/{4})=x_p(u, v)$ and $z(-{\pi}/{4})=-1$, where $\xi'\in[-\pi/4,\pi/4)$ serves as the parameter inside the circle. 

Next we solve $A, B$ and $x_p$ as functions of $u, v$.  Substituting the coordinates of the three points above into Eq.~\eqref{zxi}, one can obtain
\begin{eqnarray}
\label{A}A=-\sqrt{2}\cos\alpha>0,\\
\label{B}B=\sqrt{2}\sin\alpha>0;
\end{eqnarray}
and 
\begin{eqnarray}
\cos\eta&=& -\frac{u\cos\alpha+v\sin\alpha}{\sqrt{2}\cos\alpha},\\
\sin\eta&=&\frac{v\cos\alpha-u\sin\alpha}{\sqrt{2}\sin\alpha}.
\end{eqnarray}
 The later two equations can be combined into a quartic equation
 \begin{equation}
 \frac{v^2}{2}\beta^4+uv\beta^3+(u^2-1)\beta^2-uv\beta+\frac{v^2}{2}=0. 
\end{equation}
With the physical solution $\beta(u, v)=\tan\alpha(u, v)>0$ in mind, one can solve 
\begin{equation}
A=\sqrt{\frac{2}{\beta^2+1}}, 
\end{equation}
\begin{equation}
B=-\beta\sqrt{\frac{2}{\beta^2+1}}.
\end{equation}
\begin{eqnarray}
\cos\eta&=&-\frac{u+v\beta}{\sqrt{2}},\\
\sin\eta&=&\frac{v-u\beta}{\sqrt{2}\beta}.
\end{eqnarray}
Note that $\alpha$ needs to be in the second quadrant $(\pi/2, \pi)$ according to Eqs.~\eqref{A} and ~\eqref{B}. 

Now we are ready to write out the phase of an arbitrary point $z(\eta)=u+iv$ as
\begin{eqnarray}
\phi(u, v)=k_0\Big[b-x_p+\int_{-\frac{\pi}{4}}^\eta{\rm d}\xi'n^2(x(\xi'),y(\xi'))\Big]. 
\end{eqnarray}
knowing that it composes of the sum of phase accumulated outside and inside the lens. After a straightforward derivation~\footnote{The inverse tangent function $\arctan(x, y)$ is defined following \texttt{ArcTan} in \texttt{Wolfram Mathematica 10}, such that it gives the the right angle of $(x, y)$ among four quadrants.}
\begin{eqnarray}
x_p(u, v)&=&1-\frac{2}{\beta^2+1},\\
\label{phiuv}
\phi(u, v)&=&k_0\Big[b-1+\frac{2}{\beta^2+1}+\eta(u,v)+\frac{\pi}{4}+\nonumber\\&&\frac{\beta^2-1}{2(\beta^2+1)}\Big(1-(u+v\beta)\frac{v-u\beta}{\beta}\Big)\Big]. \\
\eta&=&\arctan\Big(-\frac{u+v\beta}{\sqrt{2}}, \frac{v-u\beta}{\sqrt{2}\beta}\Big).
\end{eqnarray}
The interval of parameter $\eta\in[-\pi/4, \pi/4)$ dictates the physical solutions of $\beta$ to be 
\begin{eqnarray}
\beta = \begin{cases}
-\frac{u+\sqrt{2-u^2-2v^2}}{2v}-\sqrt{\frac{1+v^2+u\sqrt{2-u^2-2v^2}}{2v^2}}, &v>0; \cr
-\frac{u+\sqrt{2-u^2-2v^2}}{2v}+\sqrt{\frac{1+v^2+u\sqrt{2-u^2-2v^2}}{2v^2}}, &v<0.
\end{cases}
\end{eqnarray} 
Note that $\beta$ becomes peculiar when $v=0$ and we can write the phase 
\begin{eqnarray}
\phi(u, v=0)&=&k_0\Big(b-\frac{1}{2}+\arccos\frac{u}{\sqrt{2}}-\frac{\pi}{4}\\\nonumber&-&\frac{u}{2}\sqrt{2-u^2}\Big).
\end{eqnarray}
Substituting the full phase Eq.~\eqref{phiuv} depicted in Fig.~\ref{fig:Fig4}(a), into our formula of permittivity Eq.~\eqref{epsf}
\begin{eqnarray}
\epsilon&\approx&\frac{1}{k_0^2}\Bigg[\vert\nabla \phi\vert^2
+\Big(\frac{\nabla^2\phi}{2\vert\nabla \phi\vert}\Big)^2+\nonumber\\
&&\frac{\nabla \phi}{2\vert\nabla \phi\vert^4}\cdot\bigg(\vert\nabla \phi\vert^2\nabla\nabla^2\phi-2\nabla^2\phi(\nabla \phi\cdot\nabla)\nabla \phi\bigg)\Bigg],
\end{eqnarray}
we calculate the designed profile as Fig.~\ref{fig:Fig4}(b-c) shows. The designed permittivity goes below unity near the source point $(-1, 0)$, which is due to the fact that the phase near source point is undefined [\emph{cf}. Fig.~\ref{fig:Fig4}(b)]. Despite that, it is clear that our method successfully reproduces that of L\"uneburg lens [see Fig.~\ref{fig:Fig4}(c)].


\bibliographystyle{apsrev4-1}
%


\begin{thebibliography}{31}%
\makeatletter
\providecommand \@ifxundefined [1]{%
 \@ifx{#1\undefined}
}%
\providecommand \@ifnum [1]{%
 \ifnum #1\expandafter \@firstoftwo
 \else \expandafter \@secondoftwo
 \fi
}%
\providecommand \@ifx [1]{%
 \ifx #1\expandafter \@firstoftwo
 \else \expandafter \@secondoftwo
 \fi
}%
\providecommand \natexlab [1]{#1}%
\providecommand \enquote  [1]{``#1''}%
\providecommand \bibnamefont  [1]{#1}%
\providecommand \bibfnamefont [1]{#1}%
\providecommand \citenamefont [1]{#1}%
\providecommand \href@noop [0]{\@secondoftwo}%
\providecommand \href [0]{\begingroup \@sanitize@url \@href}%
\providecommand \@href[1]{\@@startlink{#1}\@@href}%
\providecommand \@@href[1]{\endgroup#1\@@endlink}%
\providecommand \@sanitize@url [0]{\catcode `\\12\catcode `\$12\catcode
  `\&12\catcode `\#12\catcode `\^12\catcode `\_12\catcode `\%12\relax}%
\providecommand \@@startlink[1]{}%
\providecommand \@@endlink[0]{}%
\providecommand \url  [0]{\begingroup\@sanitize@url \@url }%
\providecommand \@url [1]{\endgroup\@href {#1}{\urlprefix }}%
\providecommand \urlprefix  [0]{URL }%
\providecommand \Eprint [0]{\href }%
\providecommand \doibase [0]{http://dx.doi.org/}%
\providecommand \selectlanguage [0]{\@gobble}%
\providecommand \bibinfo  [0]{\@secondoftwo}%
\providecommand \bibfield  [0]{\@secondoftwo}%
\providecommand \translation [1]{[#1]}%
\providecommand \BibitemOpen [0]{}%
\providecommand \bibitemStop [0]{}%
\providecommand \bibitemNoStop [0]{.\EOS\space}%
\providecommand \EOS [0]{\spacefactor3000\relax}%
\providecommand \BibitemShut  [1]{\csname bibitem#1\endcsname}%
\let\auto@bib@innerbib\@empty
\bibitem [{\citenamefont {Pendry}(2006)}]{Pendry2006}%
  \BibitemOpen
  \bibfield  {author} {\bibinfo {author} {\bibfnamefont {J.~B.}\ \bibnamefont
  {Pendry}},\ }\href {\doibase 10.1126/science.1125907} {\bibfield  {journal}
  {\bibinfo  {journal} {Science}\ }\textbf {\bibinfo {volume} {312}},\ \bibinfo
  {pages} {1780} (\bibinfo {year} {2006})}\BibitemShut {NoStop}%
\bibitem [{\citenamefont {Leonhardt}(2006{\natexlab{a}})}]{Leonhardt2006a}%
  \BibitemOpen
  \bibfield  {author} {\bibinfo {author} {\bibfnamefont {U.}~\bibnamefont
  {Leonhardt}},\ }\href {\doibase 10.1126/science.1126493} {\bibfield
  {journal} {\bibinfo  {journal} {Science}\ }\textbf {\bibinfo {volume}
  {312}},\ \bibinfo {pages} {1777} (\bibinfo {year}
  {2006}{\natexlab{a}})}\BibitemShut {NoStop}%
\bibitem [{\citenamefont {Leonhardt}(2006{\natexlab{b}})}]{Leonhardt2006b}%
  \BibitemOpen
  \bibfield  {author} {\bibinfo {author} {\bibfnamefont {U.}~\bibnamefont
  {Leonhardt}},\ }\href {\doibase 10.1088/1367-2630/8/7/118} {\bibfield
  {journal} {\bibinfo  {journal} {New Journal of Physics}\ }\textbf {\bibinfo
  {volume} {8}},\ \bibinfo {pages} {16} (\bibinfo {year}
  {2006}{\natexlab{b}})}\BibitemShut {NoStop}%
\bibitem [{\citenamefont {Leonhardt}\ and\ \citenamefont
  {Philbin}(2010)}]{Leonhardt2010book}%
  \BibitemOpen
  \bibfield  {author} {\bibinfo {author} {\bibfnamefont {U.}~\bibnamefont
  {Leonhardt}}\ and\ \bibinfo {author} {\bibfnamefont {T.}~\bibnamefont
  {Philbin}},\ }\href@noop {} {\emph {\bibinfo {title} {Geometry and Light :
  The Science of Invisibility}}},\ Dover Books on Physics\ (\bibinfo
  {publisher} {Dover Publications, Inc.},\ \bibinfo {address} {Mineola, N.Y.},\
  \bibinfo {year} {2010})\ p.\ \bibinfo {pages} {278}\BibitemShut {NoStop}%
\bibitem [{\citenamefont {Post}(1962)}]{Post1962}%
  \BibitemOpen
  \bibfield  {author} {\bibinfo {author} {\bibfnamefont {E.~J.}\ \bibnamefont
  {Post}},\ }\href@noop {} {\emph {\bibinfo {title} {Formal Structure of
  Electromagnetics--General Covariance and Electromagnetics}}}\ (\bibinfo
  {publisher} {North-Holland Publishing Companuy},\ \bibinfo {address}
  {Amsterdam},\ \bibinfo {year} {1962})\BibitemShut {NoStop}%
\bibitem [{\citenamefont {Schmiele}\ \emph {et~al.}(2010)\citenamefont
  {Schmiele}, \citenamefont {Varma}, \citenamefont {Rockstuhl},\ and\
  \citenamefont {Lederer}}]{Schmiele2010}%
  \BibitemOpen
  \bibfield  {author} {\bibinfo {author} {\bibfnamefont {M.}~\bibnamefont
  {Schmiele}}, \bibinfo {author} {\bibfnamefont {V.~S.}\ \bibnamefont {Varma}},
  \bibinfo {author} {\bibfnamefont {C.}~\bibnamefont {Rockstuhl}}, \ and\
  \bibinfo {author} {\bibfnamefont {F.}~\bibnamefont {Lederer}},\ }\href
  {\doibase 10.1103/PhysRevA.81.033837} {\bibfield  {journal} {\bibinfo
  {journal} {Physical Review A}\ }\textbf {\bibinfo {volume} {81}} (\bibinfo
  {year} {2010}),\ 10.1103/PhysRevA.81.033837}\BibitemShut {NoStop}%
\bibitem [{\citenamefont {Li}\ and\ \citenamefont {Pendry}(2008)}]{Li2008}%
  \BibitemOpen
  \bibfield  {author} {\bibinfo {author} {\bibfnamefont {J.}~\bibnamefont
  {Li}}\ and\ \bibinfo {author} {\bibfnamefont {J.~B.}\ \bibnamefont
  {Pendry}},\ }\href {\doibase 10.1103/PhysRevLett.101.203901} {\bibfield
  {journal} {\bibinfo  {journal} {Phys. Rev. Lett.}\ }\textbf {\bibinfo
  {volume} {101}},\ \bibinfo {pages} {203901} (\bibinfo {year}
  {2008})}\BibitemShut {NoStop}%
\bibitem [{\citenamefont {Chang}\ \emph {et~al.}(2010)\citenamefont {Chang},
  \citenamefont {Zhou}, \citenamefont {Hu},\ and\ \citenamefont
  {Hu}}]{Chang2010}%
  \BibitemOpen
  \bibfield  {author} {\bibinfo {author} {\bibfnamefont {Z.}~\bibnamefont
  {Chang}}, \bibinfo {author} {\bibfnamefont {X.}~\bibnamefont {Zhou}},
  \bibinfo {author} {\bibfnamefont {J.}~\bibnamefont {Hu}}, \ and\ \bibinfo
  {author} {\bibfnamefont {G.}~\bibnamefont {Hu}},\ }\href {\doibase
  10.1364/OE.18.006089} {\bibfield  {journal} {\bibinfo  {journal} {Opt.
  Express}\ }\textbf {\bibinfo {volume} {18}},\ \bibinfo {pages} {6089}
  (\bibinfo {year} {2010})}\BibitemShut {NoStop}%
\bibitem [{\citenamefont {Liu}\ \emph {et~al.}(2013)\citenamefont {Liu},
  \citenamefont {Gabrielli}, \citenamefont {Lipson},\ and\ \citenamefont
  {Johnson}}]{LiuD2013}%
  \BibitemOpen
  \bibfield  {author} {\bibinfo {author} {\bibfnamefont {D.}~\bibnamefont
  {Liu}}, \bibinfo {author} {\bibfnamefont {L.~H.}\ \bibnamefont {Gabrielli}},
  \bibinfo {author} {\bibfnamefont {M.}~\bibnamefont {Lipson}}, \ and\ \bibinfo
  {author} {\bibfnamefont {S.~G.}\ \bibnamefont {Johnson}},\ }\href {\doibase
  10.1364/OE.21.014223} {\bibfield  {journal} {\bibinfo  {journal} {Opt
  Express}\ }\textbf {\bibinfo {volume} {21}},\ \bibinfo {pages} {14223}
  (\bibinfo {year} {2013})}\BibitemShut {NoStop}%
\bibitem [{\citenamefont {Piggott}\ \emph {et~al.}(2015)\citenamefont
  {Piggott}, \citenamefont {Lu}, \citenamefont {Lagoudakis}, \citenamefont
  {Petykiewicz}, \citenamefont {Babinec},\ and\ \citenamefont
  {Vučković}}]{Piggott2015}%
  \BibitemOpen
  \bibfield  {author} {\bibinfo {author} {\bibfnamefont {A.~Y.}\ \bibnamefont
  {Piggott}}, \bibinfo {author} {\bibfnamefont {J.}~\bibnamefont {Lu}},
  \bibinfo {author} {\bibfnamefont {K.~G.}\ \bibnamefont {Lagoudakis}},
  \bibinfo {author} {\bibfnamefont {J.}~\bibnamefont {Petykiewicz}}, \bibinfo
  {author} {\bibfnamefont {T.~M.}\ \bibnamefont {Babinec}}, \ and\ \bibinfo
  {author} {\bibfnamefont {J.}~\bibnamefont {Vučković}},\ }\href {\doibase
  10.1038/nphoton.2015.69
  http://www.nature.com/nphoton/journal/v9/n6/abs/nphoton.2015.69.html#supplementary-information}
  {\bibfield  {journal} {\bibinfo  {journal} {Nat Photon}\ }\textbf {\bibinfo
  {volume} {9}},\ \bibinfo {pages} {374} (\bibinfo {year} {2015})}\BibitemShut
  {NoStop}%
\bibitem [{\citenamefont {Ginis}\ \emph {et~al.}(2016)\citenamefont {Ginis},
  \citenamefont {Tassin}, \citenamefont {Koschny},\ and\ \citenamefont
  {Soukoulis}}]{Ginis2016}%
  \BibitemOpen
  \bibfield  {author} {\bibinfo {author} {\bibfnamefont {V.}~\bibnamefont
  {Ginis}}, \bibinfo {author} {\bibfnamefont {P.}~\bibnamefont {Tassin}},
  \bibinfo {author} {\bibfnamefont {T.}~\bibnamefont {Koschny}}, \ and\
  \bibinfo {author} {\bibfnamefont {C.~M.}\ \bibnamefont {Soukoulis}},\ }\href
  {\doibase 10.1063/1.4939979} {\bibfield  {journal} {\bibinfo  {journal}
  {Applied Physics Letters}\ }\textbf {\bibinfo {volume} {108}},\ \bibinfo
  {pages} {031601} (\bibinfo {year} {2016})}\BibitemShut {NoStop}%
\bibitem [{\citenamefont {Leonhardt}\ and\ \citenamefont
  {Tyc}(2009)}]{Leonhardt2009c}%
  \BibitemOpen
  \bibfield  {author} {\bibinfo {author} {\bibfnamefont {U.}~\bibnamefont
  {Leonhardt}}\ and\ \bibinfo {author} {\bibfnamefont {T.}~\bibnamefont
  {Tyc}},\ }\href {\doibase 10.1126/science.1166332} {\bibfield  {journal}
  {\bibinfo  {journal} {Science}\ }\textbf {\bibinfo {volume} {323}},\ \bibinfo
  {pages} {110} (\bibinfo {year} {2009})}\BibitemShut {NoStop}%
\bibitem [{\citenamefont {Dalarsson}\ and\ \citenamefont
  {Tassin}(2009)}]{Dalarsson2009}%
  \BibitemOpen
  \bibfield  {author} {\bibinfo {author} {\bibfnamefont {M.}~\bibnamefont
  {Dalarsson}}\ and\ \bibinfo {author} {\bibfnamefont {P.}~\bibnamefont
  {Tassin}},\ }\href {\doibase 10.1364/OE.17.006747} {\bibfield  {journal}
  {\bibinfo  {journal} {Opt. Express}\ }\textbf {\bibinfo {volume} {17}},\
  \bibinfo {pages} {6747} (\bibinfo {year} {2009})}\BibitemShut {NoStop}%
\bibitem [{\citenamefont {Vial}\ \emph {et~al.}(2016)\citenamefont {Vial},
  \citenamefont {Liu}, \citenamefont {Horsley}, \citenamefont {Philbin},\ and\
  \citenamefont {Hao}}]{Ben2016b}%
  \BibitemOpen
  \bibfield  {author} {\bibinfo {author} {\bibfnamefont {B.}~\bibnamefont
  {Vial}}, \bibinfo {author} {\bibfnamefont {Y.}~\bibnamefont {Liu}}, \bibinfo
  {author} {\bibfnamefont {S.~A.~R.}\ \bibnamefont {Horsley}}, \bibinfo
  {author} {\bibfnamefont {T.~G.}\ \bibnamefont {Philbin}}, \ and\ \bibinfo
  {author} {\bibfnamefont {Y.}~\bibnamefont {Hao}},\ }\href@noop {} {\bibfield
  {journal} {\bibinfo  {journal} {Physical Review B}\ }\textbf {\bibinfo
  {volume} {94}},\ \bibinfo {pages} {245119} (\bibinfo {year} {2016})},\
  \bibinfo {note} {https://arxiv.org/abs/1608.05642v2}\BibitemShut {NoStop}%
\bibitem [{\citenamefont {Philbin}(2014)}]{Philbin2014}%
  \BibitemOpen
  \bibfield  {author} {\bibinfo {author} {\bibfnamefont {T.~G.}\ \bibnamefont
  {Philbin}},\ }\href {\doibase http://arxiv.org/abs/1402.2811} {\bibfield
  {journal} {\bibinfo  {journal} {Journal of Modern Optics}\ }\textbf {\bibinfo
  {volume} {61}},\ \bibinfo {pages} {552} (\bibinfo {year} {2014})}\BibitemShut
  {NoStop}%
\bibitem [{\citenamefont {Kim}\ and\ \citenamefont {Park}(2013)}]{Kim2013}%
  \BibitemOpen
  \bibfield  {author} {\bibinfo {author} {\bibfnamefont {K.~H.}\ \bibnamefont
  {Kim}}\ and\ \bibinfo {author} {\bibfnamefont {Q.~H.}\ \bibnamefont {Park}},\
  }\href {\doibase 10.1038/srep01062} {\bibfield  {journal} {\bibinfo
  {journal} {Sci Rep}\ }\textbf {\bibinfo {volume} {3}},\ \bibinfo {pages}
  {1062} (\bibinfo {year} {2013})}\BibitemShut {NoStop}%
\bibitem [{Note1()}]{Note1}%
  \BibitemOpen
  \bibinfo {note} {Taking $E_z=F(x,y)\protect \qopname \relax o{exp}(i k_0 x)$
  with $F\rightarrow 1$ at infinity and plugging this into the wave equation of
  no source~\protect \textup {\hbox {\mathsurround \z@ \protect \normalfont
  (\ignorespaces \ref {Ez}\unskip \@@italiccorr )}}, one reads off the
  corresponding permittivity.}\BibitemShut {Stop}%
\bibitem [{\citenamefont {Born}\ and\ \citenamefont
  {Wolf}(2009)}]{BornWolf1999}%
  \BibitemOpen
  \bibfield  {author} {\bibinfo {author} {\bibfnamefont {M.}~\bibnamefont
  {Born}}\ and\ \bibinfo {author} {\bibfnamefont {E.}~\bibnamefont {Wolf}},\
  }\href@noop {} {\enquote {\bibinfo {title} {Principles of optics},}\ }
  (\bibinfo {year} {2009})\BibitemShut {NoStop}%
\bibitem [{Note2()}]{Note2}%
  \BibitemOpen
  \bibinfo {note} {To be compatible with the model setup in \protect \texttt
  {COMSOL} simulation, we choose the engineering time harmonic convection
  $e^{i\omega t}$.}\BibitemShut {Stop}%
\bibitem [{\citenamefont {Liu}(2014)}]{YangjieThesis}%
  \BibitemOpen
  \bibfield  {author} {\bibinfo {author} {\bibfnamefont {Y.}~\bibnamefont
  {Liu}},\ }\emph {\bibinfo {title} {Time-Varying Emission of Electrons and Its
  Relevant {EM} Phenomenon}},\ \href {\doibase
  https://repository.ntu.edu.sg/handle/10356/60623} {Ph.D. thesis} (\bibinfo
  {year} {2014})\BibitemShut {NoStop}%
\bibitem [{Note3()}]{Note3}%
  \BibitemOpen
  \bibinfo {note} {The general situation when the source sits in non-trivial
  region where material parameters are not unity, is far more complicated and
  beyond scope of this article.}\BibitemShut {Stop}%
\bibitem [{Note4()}]{Note4}%
  \BibitemOpen
  \bibinfo {note} {Due to our condition that the current source sits in trivial
  region, we cannot control the wave amplitude in the source
  region.}\BibitemShut {Stop}%
\bibitem [{Note5()}]{Note5}%
  \BibitemOpen
  \bibinfo {note} {Note that the variable $K$ as the factor before the
  perpendicular vector $\protect \mathbf {p}$ is actually \protect \emph {not}
  arbitrary as it might appear at the first sight. The reason not being so is
  two-fold: 1) the requirement that material parameters reduce to unity
  prevents us from exploiting the freedom; 2) moreover, $K(x, y)$ is restrained
  in order to compliment the former term on the right side of Eq.~\protect
  \textup {\hbox {\mathsurround \z@ \protect \normalfont (\ignorespaces \ref
  {integrability}\unskip \@@italiccorr )}} into a gradient. Therefore it
  \protect \emph {cannot} be arbitrary function of $x$ and $y$. Thus
  Eq.~\protect \textup {\hbox {\mathsurround \z@ \protect \normalfont
  (\ignorespaces \ref {integrability}\unskip \@@italiccorr )}} is an
  integrability problem to solve. Once $F(A^2, \mu )$ is known, one can design
  the wave amplitude $A$ and the permeability becomes known.}\BibitemShut
  {Stop}%
\bibitem [{\citenamefont {Tyc}\ \emph {et~al.}(2014)\citenamefont {Tyc},
  \citenamefont {Chen}, \citenamefont {Danner},\ and\ \citenamefont
  {Xu}}]{Tyc2014b}%
  \BibitemOpen
  \bibfield  {author} {\bibinfo {author} {\bibfnamefont {T.}~\bibnamefont
  {Tyc}}, \bibinfo {author} {\bibfnamefont {H.}~\bibnamefont {Chen}}, \bibinfo
  {author} {\bibfnamefont {A.}~\bibnamefont {Danner}}, \ and\ \bibinfo {author}
  {\bibfnamefont {Y.}~\bibnamefont {Xu}},\ }\href {\doibase
  10.1103/PhysRevA.90.053829} {\bibfield  {journal} {\bibinfo  {journal}
  {Physical Review A}\ }\textbf {\bibinfo {volume} {90}} (\bibinfo {year}
  {2014}),\ 10.1103/PhysRevA.90.053829}\BibitemShut {NoStop}%
\bibitem [{Note6()}]{Note6}%
  \BibitemOpen
  \bibinfo {note} {We choose only x dependence for simplicity.}\BibitemShut
  {Stop}%
\bibitem [{Note7()}]{Note7}%
  \BibitemOpen
  \bibinfo {note} {Reciprocality of wave equation implies that one could also
  choose the output to be a cylindrical wave. However, this will give a wave
  picture of inevitable scattering (not shown in the paper).}\BibitemShut
  {Stop}%
\bibitem [{\citenamefont {Gallina}\ \emph {et~al.}(2010)\citenamefont
  {Gallina}, \citenamefont {Castaldi}, \citenamefont {Galdi}, \citenamefont
  {Alù},\ and\ \citenamefont {Engheta}}]{Gallina2010}%
  \BibitemOpen
  \bibfield  {author} {\bibinfo {author} {\bibfnamefont {I.}~\bibnamefont
  {Gallina}}, \bibinfo {author} {\bibfnamefont {G.}~\bibnamefont {Castaldi}},
  \bibinfo {author} {\bibfnamefont {V.}~\bibnamefont {Galdi}}, \bibinfo
  {author} {\bibfnamefont {A.}~\bibnamefont {Alù}}, \ and\ \bibinfo {author}
  {\bibfnamefont {N.}~\bibnamefont {Engheta}},\ }\href {\doibase
  10.1103/PhysRevB.81.125124} {\bibfield  {journal} {\bibinfo  {journal}
  {Physical Review B}\ }\textbf {\bibinfo {volume} {81}},\ \bibinfo {pages}
  {125124} (\bibinfo {year} {2010})}\BibitemShut {NoStop}%
\bibitem [{Note8()}]{Note8}%
  \BibitemOpen
  \bibinfo {note} {See Eq.~(29.4) of \cite {Leonhardt2010book}}\BibitemShut
  {NoStop}%
\bibitem [{Note9()}]{Note9}%
  \BibitemOpen
  \bibinfo {note} {See \cite {Potvin2015} for a mathematical approximation in
  3D case.}\BibitemShut {Stop}%
\bibitem [{Note10()}]{Note10}%
  \BibitemOpen
  \bibinfo {note} {The inverse tangent function $\protect \qopname \relax
  o{arctan}(x, y)$ is defined following \protect \texttt {ArcTan} in \protect
  \texttt {Wolfram Mathematica 10}, such that it gives the the right angle of
  $(x, y)$ among four quadrants.}\BibitemShut {Stop}%
\bibitem [{\citenamefont {Potvin}(2015)}]{Potvin2015}%
  \BibitemOpen
  \bibfield  {author} {\bibinfo {author} {\bibfnamefont {G.}~\bibnamefont
  {Potvin}},\ }\href {\doibase 10.1364/JOSAA.32.001848} {\bibfield  {journal}
  {\bibinfo  {journal} {Journal of the Optical Society of America. A, Optics,
  image science, and vision}\ }\textbf {\bibinfo {volume} {32}},\ \bibinfo
  {pages} {1848} (\bibinfo {year} {2015})}\BibitemShut {NoStop}%
\end{thebibliography}

\end{document}